\documentclass[12pt]{iopart}
\usepackage{graphicx,epsf}
\usepackage{dcolumn}
\usepackage{bm}

\begin{document}
\title{Upper limits on neutrino 
masses from the 2dFGRS and WMAP: the role of priors}
\author{\O ystein Elgar\o y\dag\  and Ofer Lahav\ddag}
\address{\dag\ NORDITA, Blegdamsvej 17, DK-2100 Copenhagen, Denmark}
\address{\ddag\ Institute of Astronomy, University of Cambridge, Madingley 
Road, Cambridge CB3 0HA, UK}
\eads{\mailto{oelgaroy@nordita.dk}, \mailto{lahav@ast.cam.ac.uk}}

\begin{abstract}
Solar, atmospheric, and reactor neutrino experiments have confirmed
neutrino oscillations, implying that neutrinos have non-zero mass,
but without pinning down their absolute masses.
While it is established that the effect of neutrinos on the evolution of cosmic
structure is small, the upper limits derived from large-scale structure 
could help significantly to constrain the absolute scale of the 
neutrino masses. 
In a recent paper the 2dF Galaxy Redshift Survey (2dFGRS) team
provided an upper limit $m_{\nu,\rm tot} < 2.2\;{\rm eV}$ ,
i.e. approximately 0.7 eV for each of the three neutrino flavours, or
phrased in terms of their contribution to the matter density,  
$\Omega_{\nu} / \Omega_{\rm m} < 0.16$. 
Here we discuss this analysis in greater detail, considering
issues of assumed `priors' like the matter density $\Omega_{\rm m}$ and  
the bias of the galaxy distribution with respect to the dark matter
distribution.  As the suppression of the power
spectrum depends on the ratio $\Omega_{\nu}/\Omega_{m}$, we find
that the out-of-fashion Mixed Dark Matter model, with $\Omega_{\nu}=0.2$,
$\Omega_{m}=1 $ and no cosmological constant, fits both the 2dFGRS power
spectrum and the CMB data reasonably well, but only for a Hubble constant  
$H_0 < 50\;{\rm km}\,{\rm s}^{-1}\,{\rm Mpc}^{-1}$. As a consequence, 
excluding low values of the Hubble constant, e.g. with the HST Key Project, 
is important in order to get a strong upper limit on the neutrino masses.    
We also comment on the improved limit obtained by the WMAP team, and point out that the 
main neutrino
signature comes from the 2dFGRS and the Lyman $\alpha$ forest.
\end{abstract}

\maketitle

\section{Introduction}

The wealth of new data from e.g. the cosmic microwave background 
(CMB) and large-scale structure (LSS) in the last few years  
indicate that we live in a flat Universe where  
$\sim 70\;\%$ of the mass-energy density is in the form of dark energy, 
with matter making up the remaining 30 \% .   
The WMAP data combined with 
other large-scale structure data \cite{wmap} gives impressive 
support to this picture. 
Furthermore, the baryons contribute only a fraction $f_{\rm b} 
= \Omega_{\rm b}/\Omega_{\rm m} \sim 0.15$ ($\Omega_{\rm b}$ and 
$\Omega_{\rm m}$ are, respectively, the contribution of baryons and 
of all matter to the total density in units of the critical density 
$\rho_c =  3H_0^2 / 8\pi G = 1.879\times 10^{-29}h^2\;{\rm g}\,{\rm cm}
^{-3}$, where $H_0 = 100 h \;{\rm km}\,{\rm s}^{-1}\,{\rm Mpc}^{-1}$ is 
the present value of the Hubble parameter) of this, so that 
most of the matter is dark. 
The exact nature of the dark matter in the Universe is still 
unknown.  Relic neutrinos are abundant in the 
Universe, and from the observations of oscillations of 
solar and atmospheric neutrinos we know that neutrinos have 
a mass \cite{sage,sno,macro,gno,homestake,superk,gallex}
and will make 
up a fraction of the dark matter.  However, the oscillation experiments 
can only measure differences in the squared masses of the neutrinos, 
and not the absolute mass scale, so they cannot tell us 
how much of the dark matter is in neutrinos.  
From general arguments on structure formation in the Universe we know 
that most of the dark matter has to be cold, i.e. non-relativistic 
when it decoupled 
from the thermal background.  Neutrinos with masses on the 
eV scale or below will be a hot component of the dark matter.  
If they were the dominant dark-matter component, structure in the 
Universe would have formed first at large scales, and smaller structures 
would form by fragmentation (the `top-down' scenario).  
However, the combined observational 
and theoretical knowledge about large-scale structure gives strong evidence 
for the `bottom-up' picture of structure formation, i.e. structure formed 
first at small scales.  Hence, neutrinos cannot make up 
all of the dark matter (see \cite{primack} for a review).       
Neutrino experiments give some constraints on how much of the dark 
matter can be in the form of neutrinos.
Studies of the energy spectrum in 
tritium decay \cite{mainz} provide an upper limit on the electron 
neutrino mass of 2.2 eV (95 \% confidence limit).
For the effective neutrino mass scale involved in neutrinoless 
double beta decay an upper limit of 0.34 eV (90 \% confidence) 
has been inferred \cite{klapdor1} 
but then under the assumptions that neutrinos are Majorana particles 
(i.e. their own antiparticles), and the translation from 
the effective neutrino mass scale in neutrinoless double beta decay to 
neutrino mass eigenvalues requires assumptions about the neutrino mass 
hierarchy and the CP phases in the neutrino mixing matrix.  

From cosmology, 
an analysis the 2dF Galaxy Redshift Survey \cite{elgar} 
found $m_{\nu,\rm tot} < 2.2\;{\rm eV}$ as an upper limit on the 
sum of the (degenerate) mass eigenvalues.  In \cite{lewis} the 
2dFGRS was combined with pre-WMAP CMB data to give an upper limit 
$m_{\nu,\rm tot} < 0.9\;{\rm eV}$.  
The WMAP team \cite{wmap} improved this result to 
$m_{\nu,\rm tot} < 0.71\;{\rm eV}$ 
(95 \% confidence) from a combination of WMAP, ACBAR \cite{acbar}, CBI 
\cite{cbi} for the CMB, and 2dFGRS and the power spectrum inferred from 
Lyman $\alpha$ forest \cite{lyalpha1,gnedin} for large scale structure, 
a factor of roughly three better than the 2dFGRS limit (not an order 
of magnitude as stated in \cite{wmap}).  This limit is comparable 
to what the pioneering study in \cite{hutegmark} predicted would be 
possible with the Sloan Digital Sky Survey \cite{SDSS} and 
WMAP \cite{hutegmark}, and have implications for neutrino 
oscillation experiments as it seems to call into question the 
Liquid Scintillator Neutrino Detector (LSND) result \cite{LSND}, 
where the mass-square 
difference involved was $\sim 1\;{\rm eV}^2$ \cite{pierce,bhatta}.  
(However, as pointed out in \cite{steen3} it is premature to say 
that cosmology rules out the LSND results.)    
Note that neutrinos with eV masses are basically indistinguishable 
from cold dark 
matter at the epoch of last scattering, and therefore they have little 
effect on the CMB.  
The important role of the WMAP data in the cosmological neutrino mass 
limit is to break degeneracies in the parameter space 
that will otherwise limit the 
ability to constrain neutrino masses from the large-scale structure data.

In this paper we discuss in detail the recent cosmological 
neutrino mass limits, concentrating on the 2dFGRS and the 
WMAP + 2dFGRS limits and the various parameter degeneracies 
involved in the analysis.  In particular we discuss the role of 
the bias of the galaxy distribution with respect to the mass 
distribution, non-linear effects, and the necessity of using 
independent information about cosmological parameters like 
$\Omega_{\rm m}$ and $h$ (`priors').  
We will throughout this paper work within 
the context of flat $\Lambda{\rm CDM}$ models, 
which are favoured by a wealth of 
observational data \cite{wmap,efstathiou,lewis}, however we also 
comment on Mixed Dark Matter (MDM) models in discussion of the 
analysis of the 2dFGRS data.  In fact we find that an MDM 
model can still provide a reasonable fit to the 2dFGRS 
and WMAP data, although with a low Hubble constant, so that 
external constraints on the Hubble constant are important in order 
to get a strong upper limit on the neutrino masses. 

The structure of this paper is as follows: in section 2 we give a 
brief overview of how neutrinos affect structure formation in the 
Universe.  
In section 3 we consider galaxy redshift surveys as a probe of 
neutrino masses, starting with a brief summary of the analysis 
in \cite{elgar}.  Since the exact relationship between the 
distribution of the galaxies and that of the dark matter is 
unknown, we discuss different ways of taking this uncertainty 
into account.   We also discuss the role of priors on parameters degenerate 
with massive neutrinos. 
In section 4 we give a brief overview of other cosmological probes 
of neutrino masses before we summarize and conclude in section 5.

\section{Massive neutrinos and structure formation}

The relic abundance of neutrinos in the Universe today is 
straightforwardly found from the fact that they continue to 
follow the Fermi-Dirac distribution after freeze-out, and their 
temperature is given in terms of the CMB temperature $T_{\rm CMB}$ today as 
$T_\nu = (4/11)^{1/3}T_{\rm CMB}$, 
\begin{equation}
n_\nu = \frac{6\zeta(3)}{11\pi^2}T_{\rm CMB}^3, 
\label{eq:relicabund}
\end{equation}
where $\zeta(3)\approx 1.202$, 
which gives $n_\nu \approx 112\;{\rm cm}^{-3}$ at present. 
Neutrinos are so light that they were ultra-relativistic at 
freeze-out.  Their present contribution to the mass density can 
therefore be found by multiplying $n_\nu$ with the total mass 
of the neutrinos $m_{\nu,\rm tot}$, giving 
\begin{equation}
\Omega_\nu h^2 = \frac{m_{\nu,\rm tot}}{94\;{\rm eV}},
\label{eq:omeganu}
\end{equation}
for $T_{\rm CMB}=2.728\;{\rm K}$.  Several effects could  
modify this simple relation.  If any of the neutrino chemical 
potentials were initially non-zero, or there were a sizable  
neutrino-antineutrino asymmetry, this would increase 
the energy density in neutrinos and give an additional contribution 
to the radiation energy density.  However, from Big Bang Nucleosynthesis 
(BBN) one gets a very tight limit on the electron neutrino chemical potential, 
since the electron neutrino is directly involved in the processes that 
set the neutron-to-proton ratio.  Also, within the standard three-neutrino 
framework one can extend this limit to the other flavours as well.   
The recent results of the KamLAND experiment \cite{eguchi} 
confirmed the Large Mixing Angle (LMA) solution for the solar 
neutrino oscillations, and combined with the atmospheric data indicating 
maximal mixing in this sector, it has been shown that flavour equilibrium 
is established between all three neutrino species before the epoch 
of BBN \cite{dolgov,ywong,kevork}, so that the BBN constraint on the 
electron neutrino asymmetry 
applies to all flavours, which in turn implies that the lepton 
asymmetry cannot be large enough to give a significant contribution 
to the radiation energy density.  Recent analyses of WMAP and 2dFGRS 
data give independent, although not quite as strong, evidence for 
small lepton asymmetries \cite{steen3,elena}.
Within the standard picture, 
equation (\ref{eq:relicabund}) should be accurate, and therefore 
any constraint on the cosmic mass density of neutrinos should translate 
straightforwardly into a constraint on the total neutrino mass, 
according to equation (\ref{eq:omeganu}).  
If a fourth, light `sterile' neutrino exists, sterile-active oscillations 
would modify this conclusion.  No sterile neutrinos are required to 
explain the solar and atmospheric neutrino oscillation data 
\cite{pakvasa}, and the only hint so far comes from the 
possible detection of $\overline{\nu}_\mu \rightarrow \overline{\nu}_e$ 
oscillations with a small mixing angle and a mass-square difference 
$\sim 1\;{\rm eV}^2$ at the LSND  
\cite{LSND}.   Since there are only two independent mass-squared 
differences in the standard three-neutrino scenario, and they are 
orders of magnitude smaller, this hints at the existence of a 
fourth, light sterile neutrino.  However, as said, this has little 
support in the solar and atmospheric data.  The status 
of the LSND results will in the near future be clarified  by  
the MiniBooNE experiment \cite{mboone}.

\begin{figure}
\begin{center}
\includegraphics[width=84mm]{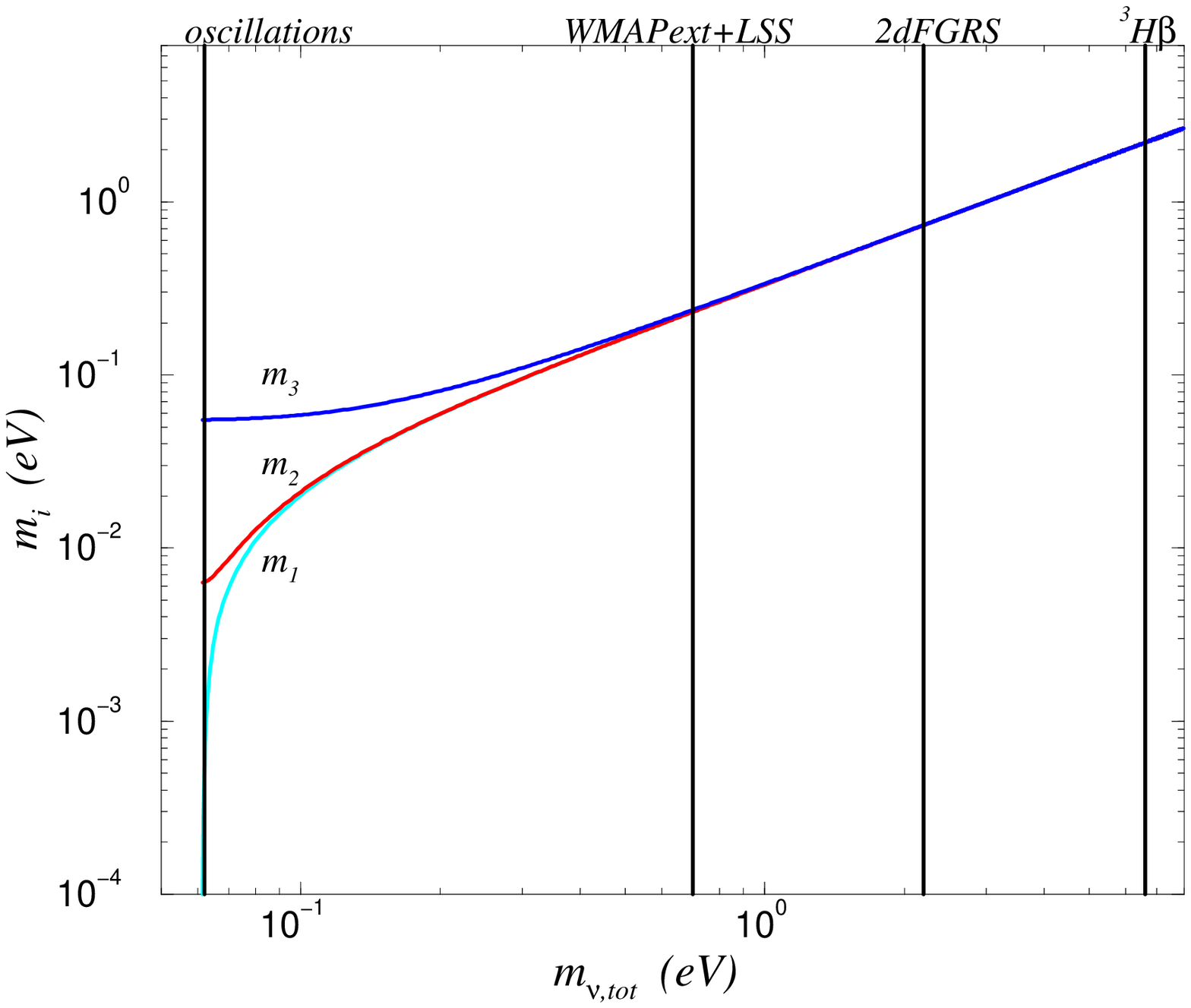}
\includegraphics[width=84mm]{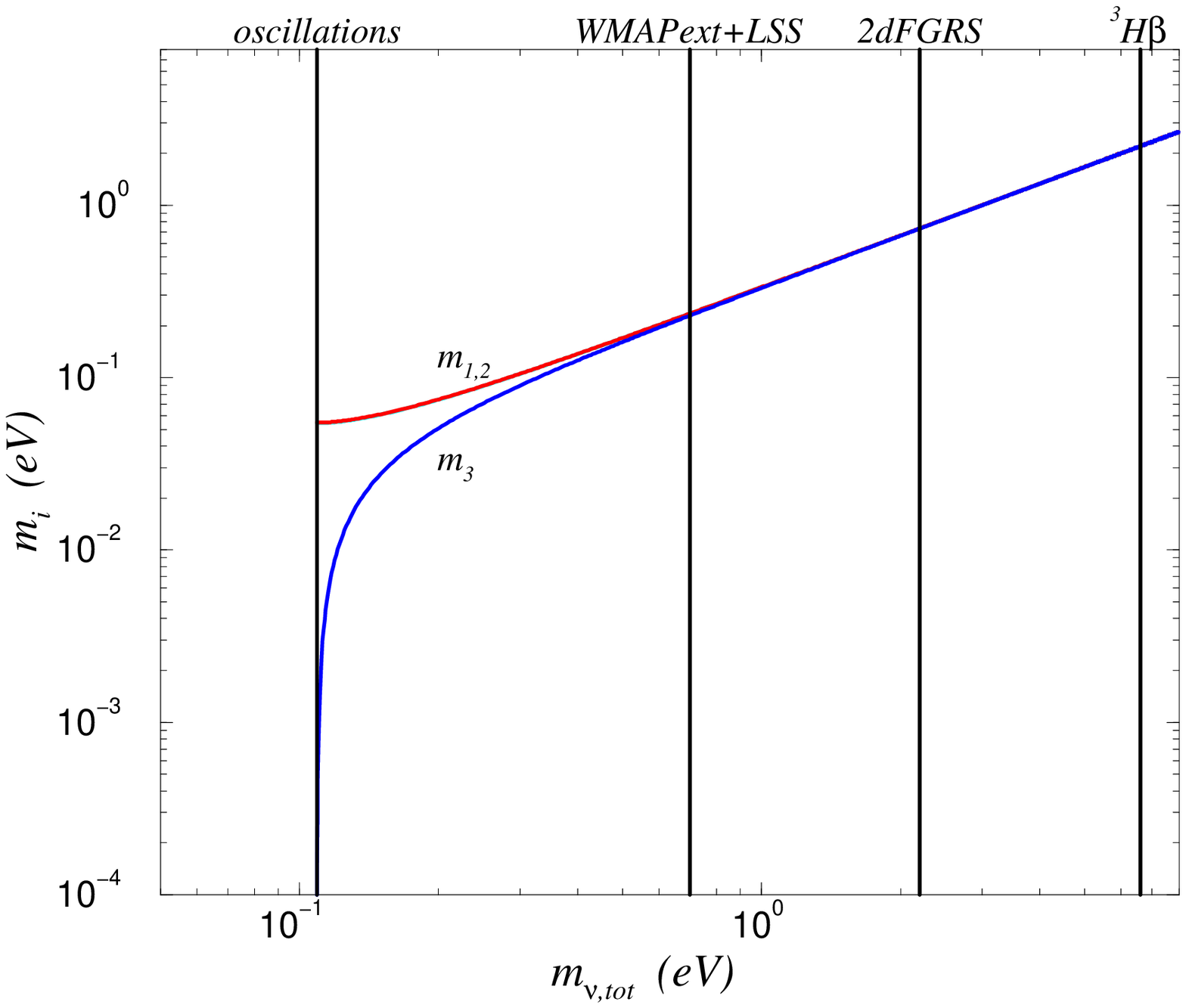}
\end{center}
\caption{Neutrino mass eigenvalues as functions of their total for 
the cases of normal (top panel) and inverted (bottom panel) hierarchies.  
The vertical line marked `oscillations' is the lower limit derived from 
the measured mass-squared differences for the two hierarchies. 
The other vertical lines are upper limits from 
WMAP+CBI+ACBAR+2dFGRS+Ly $\alpha$, 2dFGRS, and $^3{\rm H}$ $\beta$ decay.}
\label{fig:fig1}
\end{figure}

Finally, we assume that the neutrinos are 
nearly degenerate in mass.  Current cosmological observations 
are sensitive to neutrino masses  $\sim 1\;{\rm eV}$ or greater.   
Since the mass-square differences are
small, the assumption of a degenerate mass hierarchy is therefore 
justified.  This is illustrated in figure \ref{fig:fig1}, 
where we have plotted the mass eigenvalues $m_1,m_2,m_3$ as 
functions of $m_{\nu,\rm tot} = m_1 + m_2 + m_3$ for 
$\Delta m_{21}^2 = 5\times 10^{-5}\;{\rm eV}^2$ (solar) and 
$\Delta m_{32}^2 = 3 \times 10^{-3}\;{\rm eV}^2$ (atmospheric), 
for the cases of a normal hierarchy ($m_1 < m_2 < m_3$), and 
an inverted hierarchy ($m_3 < m_1 < m_2$).  As seen in the figure, 
for $m_{\nu,\rm tot} > 0.4\;{\rm eV}$ the mass eigenvalues are 
essentially degenerate. 

We will in this paper look at cosmological models 
with four components: baryons, cold dark matter, massive neutrinos, 
and a cosmological constant.  Furthermore, we restrict ourselves 
to adiabatic, linear perturbations.  The basic physics is then 
fairly simple.  A perturbation mode of a given wavelength $\lambda$ 
can grow if it is greater than the Jeans wavelength $\lambda_{\rm J}$ 
determined by the balance of gravitation and pressure, or rms velocity 
in the case of massless particles.  Above the 
Jeans scale, perturbations grow at the same rate independently of 
the scale.  Long after 
matter-radiation equality, all interesting scales are above 
$\lambda_{\rm J}$ and grow at the same rate, and in models where 
all the dark matter is cold, the  time and scale dependence of the 
power spectrum can therefore be 
separated at low redshifts.  Light, massive neutrinos can, however, move 
unhindered out regions below a certain limiting length scale, 
and will therefore tend to damp a 
density perturbation at a rate 
which depends on their rms velocity.    
The presence of massive neutrinos 
therefore introduces a new length scale, given by the size of the 
co-moving Jeans length when the neutrinos became non-relativistic.  
In terms of the comoving wavenumber, this is given by  
\begin{equation}
k_{\rm nr} = 0.015\left(\frac{m_{\nu,\rm tot}}{1\;{\rm eV}}\right)^{1/2}
\Omega_{\rm m}^{1/2}\;h\,{\rm Mpc}^{-1},
\label{eq:knr}
\end{equation}
for three equal-mass neutrinos.  The growth of Fourier modes with 
$k > k_{\rm nr}$ will be suppressed because of neutrino free-streaming.  
The free-streaming scale varies with the cosmological epoch,
and the scale and time dependence of the power spectrum cannot 
be separated, in contrast to the situation for models with cold dark 
matter only.

The transfer functions of the perturbations in the various components 
provide a convenient way of describing their evolution on different 
scales.  Using the redshift $z$ to measure time, the transfer function 
is formally defined as   
\begin{equation}
T(k,z) = \frac{\delta(k,z)}{\delta(k,z=z_*)D(z_*)}
\label{eq:tfunc}
\end{equation}
where $\delta(k,z)$ is the density perturbation with wavenumber $k$ 
at redshift $z$, and $D$ is the linear growth factor.   
The normalization redshift $z_*$ corresponds to a time long before 
the scales of interested have entered the horizon.  
The transfer function thus gives the amplitude of 
a given mode $k$ at redshift $z$ relative to its initial value, and 
is normalized so that $T(k=0,z)=1$.  The power spectrum of the matter 
fluctuations can be written as 
\begin{equation}
P_{\rm m}(k,z)= P_*(k)T^2(k,z),
\label{eq:pmatter}
\end{equation}
where $P_*(k)$ is the primordial spectrum of matter fluctuations, 
commonly assumed to be a simple power law $P_*(k)=Ak^{n}$, where 
$A$ is the amplitude and the spectral index $n$ is close to 1.  
It is also common to define power spectra for 
each component, see \cite{eisenstein} for a discussion.  
Note that the transfer functions and power spectra are independent 
of the value of the cosmological constant as long as it does not 
shift the epoch of matter-radiation equality significantly.  

\begin{figure}
\begin{center}
\includegraphics[width=84mm]{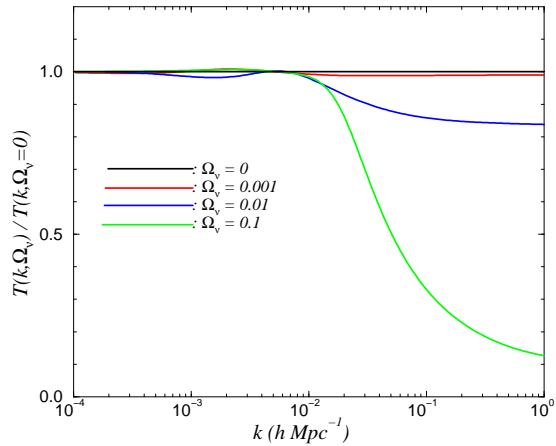}
\end{center}
\caption{Ratio of the transfer functions for various values 
of $\Omega_\nu$ to the one for $\Omega_\nu = 0$.  The other parameters 
are fixed at $\Omega_{\rm m}=0.3$, $\Omega_{\rm b}=0.04$, $h=0.7$.}
\label{fig:fig2}
\end{figure}

Accurate determination of the transfer function requires the solution 
of the coupled fluid and Boltzmann equations for the various components.  
This can be done using one of the publicly available codes, e.g. 
CMBFAST \cite{zaldarriaga} or CAMB \cite{camb}.  
Analytical approximations are also 
available, and they are very useful when one wants very quick 
computation of transfer functions.  
Accurate fitting formulas for the transfer function were derived by 
\cite{eisenstein}.  These analytic approximations are good at 
realistic baryon fractions, i.e. 0.1-0.2, with the errors typically 
smaller than 4 \%.  
In figure \ref{fig:fig2} we show the transfer functions for models 
with $\Omega_{\rm m}=0.3$, $\Omega_{\rm b}=0.04$, 
$h=0.7$ held constant, but with varying neutrino fraction.  One can 
clearly see that the small-scale suppression of power becomes more 
pronounced as the neutrino fraction $f_{\nu}\equiv \Omega_{\nu}/\Omega_
{\rm m}$ increases.  The suppression of the power spectrum on small 
scales is roughly proportional to $f_{\nu}$: 
\begin{equation}
\frac{\Delta P_{\rm m}(k)}{P_{\rm m}(k)} \sim -8f_\nu .
\label{eq:fnusuppr}
\end{equation}  

\section{Constraining the total neutrino mass with the 2\lowercase{d}FGRS}

In an earlier short paper \cite{elgar} we used the power spectrum of the galaxies 
as measured by the 2dFGRS to limit the fractional contribution 
$f_\nu = \Omega_\nu / \Omega_{\rm m}$ to the matter density of massive 
neutrinos, and on their total mass 
$m_{\nu,{\rm tot}} = 94\Omega_\nu h^2 \;{\rm eV}$.  
The present section starts with a short summary of the 2dFGRS and 
the analysis in \cite{elgar} before going into a more detailed discussion 
of the various ingredients involved in the analysis.   
In \cite{wmap} the WMAP team derived a stronger limit on the 
total neutrino mass than what was obtained from the 2dFGRS + various 
priors.  However, the 2dFGRS power spectrum played a central role 
in the WMAP neutrino mass limit. As the CMB is insensitive to 
neutrino masses in the eV range, the main role of the WMAP data 
is to provide tight constraints on parameters that may 
otherwise partly mimic the effect of massive neutrinos on the 
matter power spectrum.  Therefore our discussion of priors should 
also be of interest in understanding how the WMAP limit was obtained.  

\subsection{The 2\lowercase{d}F Galaxy Redshift Survey}
The 2dF Galaxy Redshift Survey \cite{colless} has measured the redshifts 
of more than 230 000 galaxies with a median redshift of 
$z_{\rm m} \approx 0.11$.  One of the main goals of the survey was to 
measure the galaxy power spectrum on scales up to a few hundred Mpc, 
thus filling in the gap between the small scales covered by earlier 
galaxy surveys and the largest scales where the power spectrum is constrained 
by observations of the CMB.  A sample of the size of the 2dFGRS survey allows 
large-scale structure statistics to be measured with very small random errors.
An initial estimate of the convolved, redshift-space power spectrum of the 
2dFGRS has been determined \cite{percival} based on a sample of 
140 000 redshifts. 
On scales $0.02 < k < 0.15h\;{\rm Mpc}^{-1}$ the data are robust and the 
shape of the power spectrum is not affected by redshift-space or nonlinear 
effects, though the amplitude is increased by redshift-space distortions.  
One should bear in mind that what is measured is the 
convolution of the true galaxy power spectrum with the window function 
of the survey \cite{percival,gramann}, 
\begin{equation}
P_{\rm conv}({\bf k}) \propto \int P_{\rm g}({\bf k}-{\bf q})
|W_k({\bf q})|^2 d^3q,
\label{eq:convolve1}
\end{equation}
where $W$ is the window function.  

\begin{figure}
\begin{center}
\includegraphics[width=84mm]{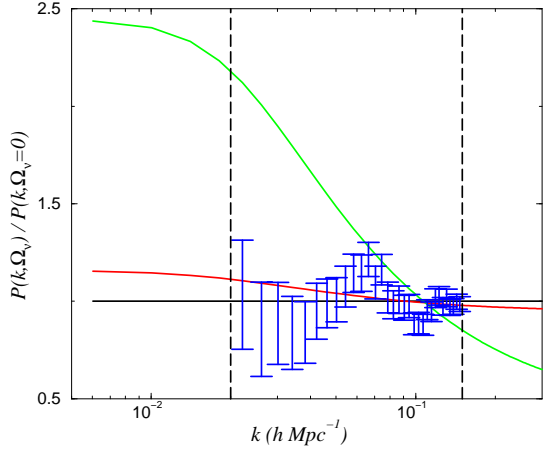}
\end{center}
\caption{Ratio of power spectra for  
$\Omega_\nu =0.01$ (bottom line) and $\Omega_\nu=0.05$ (top line)  
to the one for $\Omega_\nu = 0$ (horizontal line)   
with amplitudes fitted to the 2dFGRS power spectrum data (vertical bars) 
in redshift space.  We have fixed $\Omega_{\rm m}=0.3$, 
$h=0.7$, and $\Omega_{\rm b}h^2 = 0.02$.  The vertical dashed lines limit the 
range in $k$ used in the fits.}
\label{fig:fig3}
\end{figure}
As an illustration of the potential of the 2dFGRS to constrain neutrino 
masses, we show in figure \ref{fig:fig3} the ratio of the power spectra
for  $\Omega_\nu = 0.01$, and $0.05$ (all other
parameters are fixed at their `concordance model' values given
in the figure caption) to the power spectrum for $\Omega_\nu=0$
after they have been convolved with the survey window function, 
and their amplitudes
fitted to the 2dFGRS power spectrum data.  It is seen from figure 
\ref{fig:fig3} that the error bars on the power spectrum data points 
are correlated, as discussed in \cite{percival}, and this is taken 
into account throughout this paper by using the full covariance matrix 
of the data when computing likelihoods.  
For the 32 data points, the $\Omega_\nu=0$-model had $\chi^2=32.9$, 
$\Omega_\nu=0.01$ gives $\chi^2=33.4$,whereas the model with
$\Omega_\nu=0.05$ provides a poor fit to the data with $\chi^2=92.2$.

\subsection{Previous results from the 2dFGRS} 

In \cite{elgar} six parameters were used to describe the matter 
power spectrum: 
\begin{itemize}
\item The fraction of massive neutrinos (hot dark matter) 
$f_\nu \equiv \Omega_\nu / \Omega_{\rm m}$.  
\item The combination $\Omega_{\rm m}h$, which describes the 
shape of the cold dark matter power spectrum. 
\item The baryon fraction $f_{\rm b} \equiv \Omega_{\rm b}/\Omega_{\rm m}$.
\item The present value of the Hubble parameter $h$. 
\item The scalar spectral index of the primordial density perturbation 
spectrum, $n$.
\item The amplitude $A$ of the galaxy power spectrum. 
\end{itemize}
The amplitude is a free parameter to take into account the fact that what is 
measured is the power spectrum of galaxies, not of all the matter, 
and that it is measured in redshift space, not in real space.  
The latter effect has been shown in \cite{percival} to correspond to 
a shift in the amplitude of the spectrum.  The first effect is parametrized 
by the so-called bias parameter 
\begin{equation}
b^2(k) = \frac{P_{\rm g}(k)}{P_{\rm m}(k)}.
\label{eq:biasfac}
\end{equation}
The scale-dependence of the bias factor is not well known.  
We will discuss this in more detail in a later subsection, 
for the time being we note that one would expect the relation between the 
distribution of dark matter and luminous matter to be simple on 
large scales.  This is borne out by numerical simulations \cite{benson}, 
and two independent analyses have shown that the 2dFGRS power spectrum 
is consistent with a constant bias on the scales relevant for our 
analysis \cite{lahav,verde}.  Thus, we took the redshift space distortions 
and the bias into account by leaving the amplitude $A$ of the power 
spectrum as a free parameter, which means that we used the    
{\it shape} of the 2dFGRS power spectrum, and not its amplitude, 
to constrain $f_\nu$. 

The main effect of the massive neutrinos is to reduce the power 
on scales smaller than the neutrino free-streaming scale.  
This effect may, however, be partially masked by other effects.  
Obviously, lowering the amplitude or the scalar spectral index $n$ 
will reduce the power.  Also, the baryon 
fraction and $\Omega_{\rm m}h$ interfere with the neutrino signal.  
Therefore, constraints on  the neutrino mass from the galaxy 
power spectrum depends on the information we have about other 
parameters (`priors').  In \cite{elgar} we added 
constraints from independent cosmological probes: 
a Gaussian prior on the Hubble parameter $h=0.70\pm 0.07$, 
consistent with the results from the 
HST Hubble Key Project \cite{freedman}, 
a Gaussian prior $\Omega_{\rm b}h^2 =0.020\pm 0.002$ on the 
physical baryon density from Big Bang Nucleosynthesis \cite{burles}, 
and a Gaussian prior $n=1.0\pm 0.1$ on the scalar spectral index.      
Furthermore, we considered two different priors on $\Omega_{\rm m}$:
\begin{itemize} 
\item Under the assumption of $\Omega_{\rm m} + \Omega_\Lambda = 1$,    
a Gaussian prior $\Omega_{\rm m} = 0.28 \pm 0.14$ was obtained  
from surveys of high redshift Type Ia supernovae \cite{perlmutter,riess}. 
\item  A uniform (`top hat') prior in the range $0.1 <\Omega_{\rm m} < 0.5$.  
Given our prior on $h$, $\Omega_{\rm m} < 0.5$
ensured that the ages of the Universes in the models considered  
were greater than 12 Gyr.  
\end{itemize}
As noted earlier, the transfer function does not depend on $\Omega_\Lambda$ 
and so the assumption $\Omega_{\rm m}+\Omega_{\Lambda}=1$ enters only 
through the Supernova Type Ia prior on $\Omega_{\rm m}$. 

For each set of
parameters, we computed the theoretical matter power spectrum, and
obtained the $\chi ^2$ for the model given the 2dFGRS power spectrum.
We then calculated the joint
probability distribution function for $f_{\nu}$ and $\Gamma \equiv
\Omega_{\rm m} h$ (which represents the shape of the CDM power spectrum) by
marginalizing over $A, h$ and $f_{\rm b}$ weighted by the priors given
above.  For $A$ we used a uniform prior in the interval $0.5 < A <
10$, where $A=1$ corresponds to the normalization of the `concordance
model', discussed in \cite{lahav}.  Using instead a prior uniform in
$\log A$, or fixing $A$ at the best-fit value had virtually no effect
on the results.  
\begin{figure}
\begin{center}
\includegraphics[width=84mm]{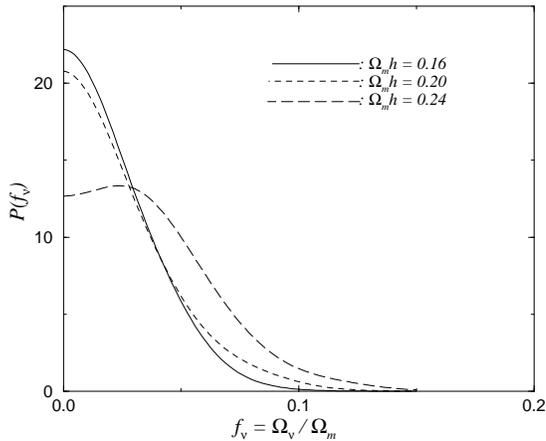}
\end{center}
\caption{Probability distributions, normalized so that the area
under each curve is equal to one, for $f_\nu$
with marginalization over the other parameters, as explained in the
text, for $N_\nu =3$ massive neutrinos and $\Omega_{\rm m}h=0.16$
(full line),
$0.20$ (dotted line), and $0.24$ (dashed line).}
\label{fig:fig4}
\end{figure}
Figure \ref{fig:fig4} shows the probability distributions for 
$f_\nu$ for three different values of $\Omega_{\rm m}h$.
Marginalizing over $\Omega_{\rm m} h$ using the uniform prior on 
$\Omega_{\rm m}$, we got an
upper limit $f_\nu < 0.13$ at 95\% confidence for $n=1$.  
Increasing $n$ increases power on small scales and leaves
more room for suppression by the massive neutrinos, and 
and upon marginalizing over the full range of $n$  
with a prior $n=1.0\pm 0.1$ we found $f_\nu < 0.16$ at 95 \% confidence.  
For $\Omega_{\rm m}h^2=0.15$ this corresponds to a total neutrino mass 
$m_{\nu,\rm tot}=2.2\;{\rm eV}$.  The results with the Supernova 
Type Ia prior on $\Omega_{\rm m}$ were identical.  
For comparison, marginalizing without any priors, the limit becomes 
$f_\nu < 0.24$.  Adding just a prior on $\Omega_{\rm m}$, we find 
$f_\nu < 0.15$, so this is clearly the most important prior.  
Marginalizing with just a prior on $h$ or on $\Omega_{\rm b}h^2$, 
the 95 \% confidence limit becomes $f_\nu < 0.20$.  
Clearly the priors play a crucial role in the analysis, and we 
will discuss their role in the next subsections. 
To facilitate the comparison with the WMAP analysis in \cite{wmap} 
we will from now on carry out our analysis in terms of the 
physical densities $\omega_{\rm i} = \Omega_{\rm i}h^2$, where 
${\rm i}={\rm m}$, $\nu$, ${\rm b}$.  

\subsection{The prior on $\omega_{\rm m}$}

\begin{figure}
\begin{center}
\includegraphics[width=84mm]{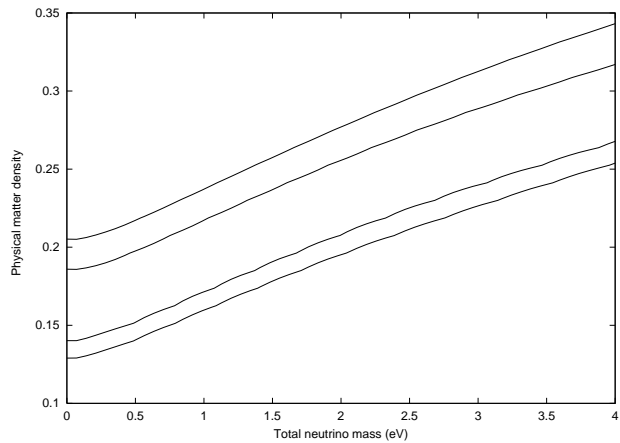}
\end{center}
\caption{Confidence contours (68 and 95 \%) from the 2dFGRS data 
alone in the plane of total neutrino mass 
$m_{\nu,\rm tot}=94\omega_\nu\;{\rm eV}$  and the physical matter 
density $\omega_{\rm m}$.  The bias parameter and $\sigma_8$ 
have been marginalized over with top-hat priors, $\omega_{\rm b}$, 
$h$, and $n$ are fixed at their best-fitting values. }
\label{fig:fig5}
\end{figure}

\begin{figure}
\begin{center}
\includegraphics[width=84mm]{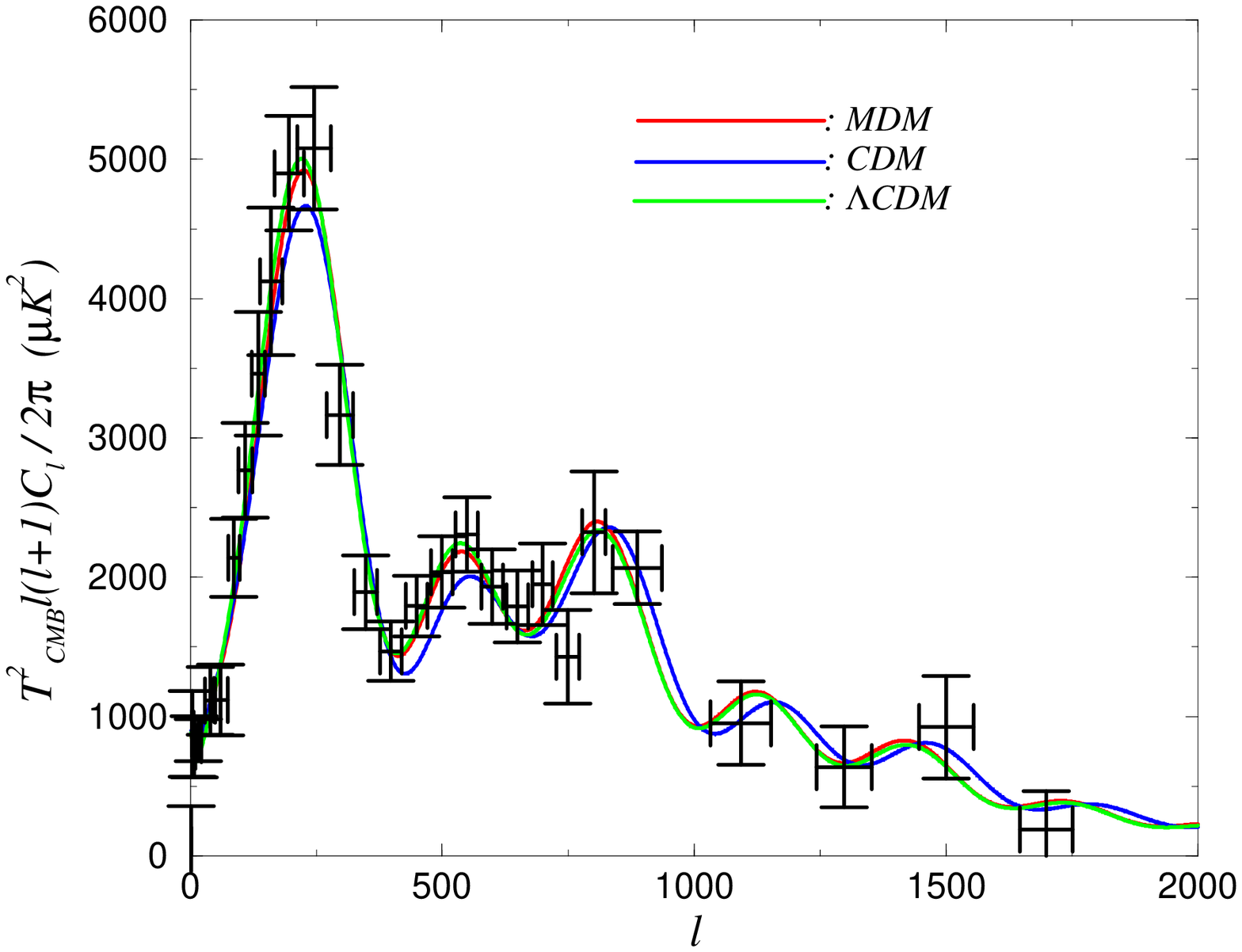}
\includegraphics[width=84mm]{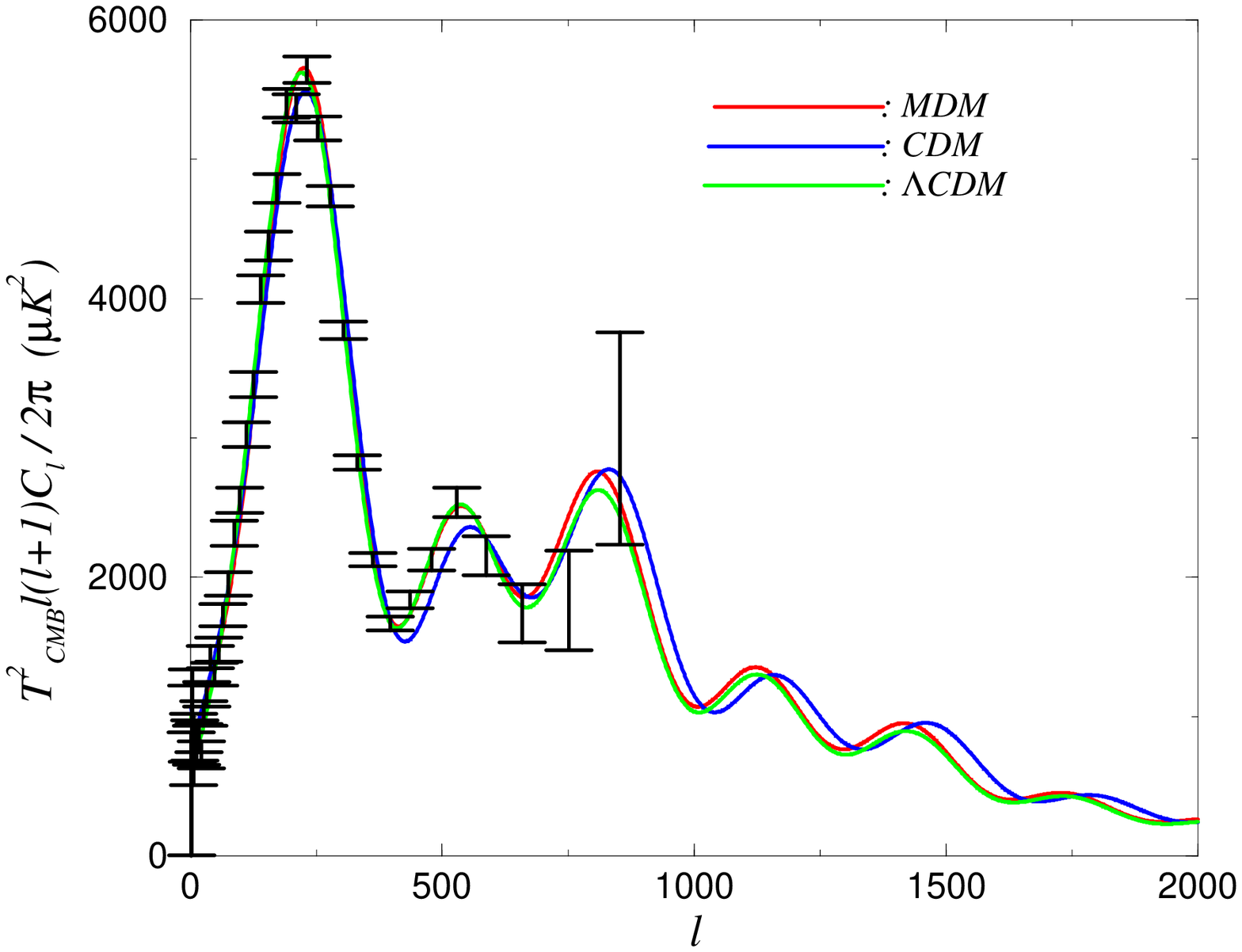}
\includegraphics[width=84mm]{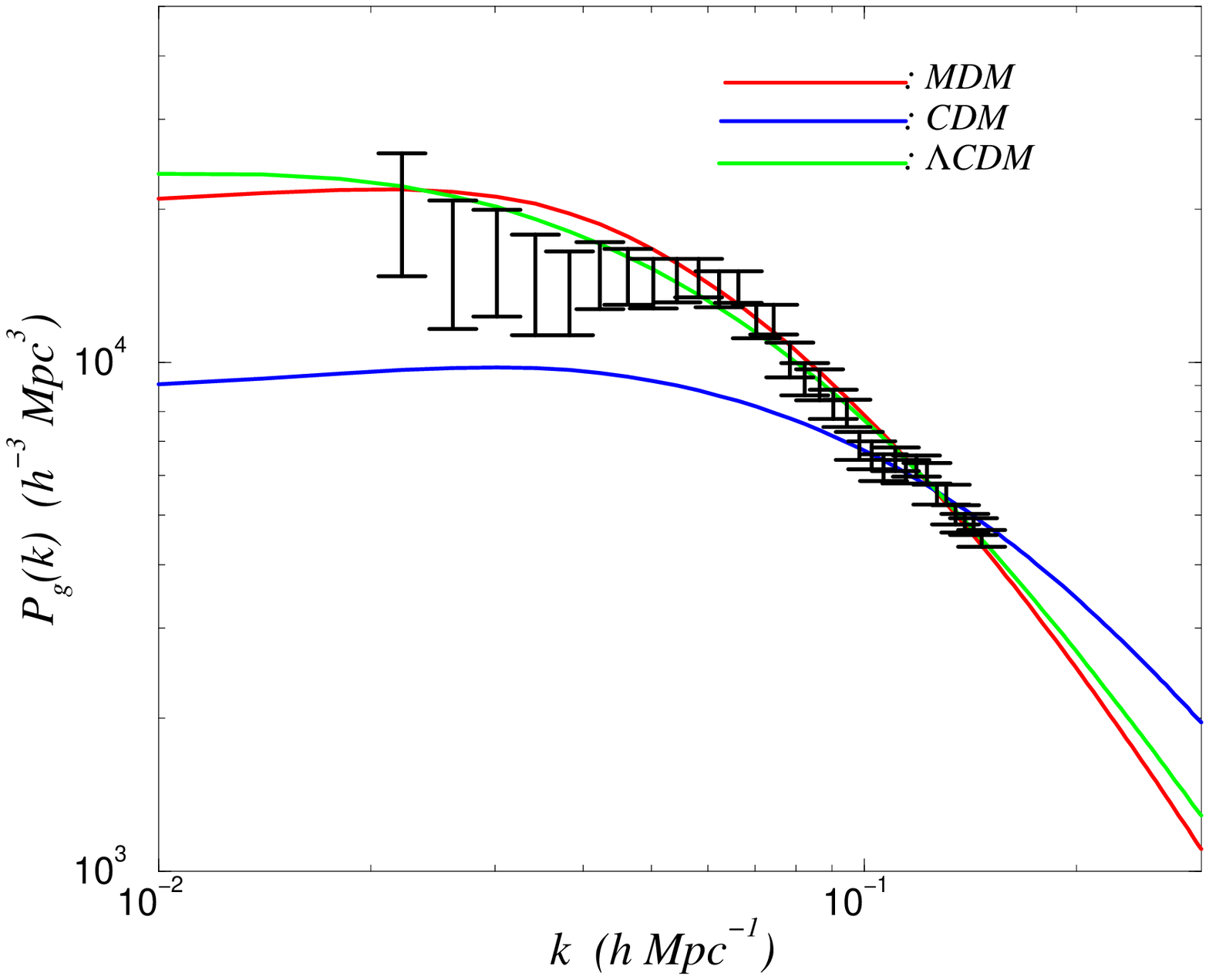}
\end{center}
\caption{MDM ($\Omega_{\rm m}=1$, $\Omega_\nu=0.2$, $h=0.45$, $n=0.95$), 
CDM ($\Omega_{\rm m}=1$, $\Omega_\nu=0$, $h=0.45$, $n=0.95$) 
and $\Lambda {\rm CDM}$ ($\Omega_{\rm m}=1$, $\Omega_\Lambda=0.7$, 
$\Omega_\nu = 0$, $h=0.7$, $n=1.0$) models (all with $\Omega_{\rm b}h^2 
= 0.024$) compared 
with data.  The upper panel shows the pre-WMAP data, the middle 
panel the WMAP data and the lower panel the 2dFGRS data along with 
the three models considered. We have normalized the models to each 
data set separately, but otherwise these are assumed models, not 
formal best fits.}
\label{fig:fig6}
\end{figure}

As noted above, the prior on the matter density is crucial, and the 
tight correlation between $m_{\nu,\rm tot}$ and $\omega_{\rm m}$
is illustrated in figure \ref{fig:fig5}.    
In fact, without a prior on $\omega_{\rm m}$ no non-trivial upper 
limit on $\omega_\nu$ is obtained (but note that one still finds 
an upper limit on $f_\nu$).  
WMAP provides a constraint $\omega_{\rm m}=0.14\pm 0.02$ for 
spatially flat $\Lambda{\rm CDM}$ models, but it is   
interesting to note that one from the CMB and 2dFGRS 
alone cannot rule out models with $\Omega_{\rm m} = \Omega_{\rm CDM} 
+ \Omega_{\rm b} + \Omega_\nu = 1$.  To illustrate this point, 
we consider the following three models, all with $\omega_{\rm b}
=0.024$:
\begin{enumerate}
\item A Mixed Dark Matter (MDM) model with $\Omega_{\rm m} =1$, 
 $\Omega_\nu = 0.2$, $h=0.45$, and $n=0.95$.  The neutrino mass 
 fraction is thus $f_\nu = 0.2$.  
\item A $\Lambda{\rm CDM}$ model with $\Omega_{\rm m} =0.3$, 
 $\Omega_\nu=0$, $h=0.7$, and $n=1.0$.
\item A pure CDM model with $\Omega_{\rm m}=1$, 
 $\Omega_\nu=0$, $h=0.45$, 
  and $n=0.95$. 
\end{enumerate}
For the pre-WMAP CMB, we use the recent compilation in \cite{wang}.  
Model 1 has $\chi^2/{\rm datum}=1.14$ for the CMB (28 points) and 
$\chi^2/{\rm datum}=1.13$ for the 32 2dFGRS power spectrum data points.  
Model 2 has $\chi^2/{\rm datum}=1.11$ for the CMB, $\chi^2=1.03$ for 
the 2dFGRS.  The WMAP data discriminate better between these two models, 
Model 1 having $\chi^2/{\rm datum}=1.15$ (899 points) and model 2 
$\chi^2/{\rm datum} = 1.08$,  but we see from figure \ref{fig:fig6} that 
the models are look reasonable and note that we have not carried out 
any systematic search for a best-fitting model of the three types.   
Thus, the first two models seem to offer acceptable descriptions of 
the CMB and 2dFGRS data, and from these data alone MDM is still a 
viable alternative to the `concordance' $\Lambda{\rm CDM}$.  So why did 
several pre-WMAP studies find that the CMB and 2dFGRS prefer a 
low matter density and a 
cosmological constant ?  This is because they  considered 
(very reasonably) neutrinos to be essentially massless.  Model 3 
illustrates the point: it is a pure CDM model with massless neutrinos.  
It gives a reasonable description of  the CMB data, but has 
$\chi^2=67$ for the 2dFGRS data points, and hence it will be 
disfavoured in a joint analysis.  The CMB cannot 
distinguish between eV-mass neutrinos and cold dark matter, and hence 
model 1 and 3 provide comparable descriptions of the CMB data, but the 
galaxy power spectrum does distinguish between the two, and massive 
neutrinos provides the necessary reduction in small-scale power to 
fit the data points.  Of course, one 
cannot look at the CMB and 2dFGRS data alone, and it is not our 
intention to `resuurect'  the MDM model: it  
has problems with the evolution of cluster abundances with 
redshift \cite{lukash}, needs 
a low value of the Hubble parameter, and $\Omega_{\rm m}=1$ is clearly 
at variance with independent measurements from e.g. the baryon 
fraction in clusters \cite{pirin}, but we wish to make the 
two following points:
\begin{itemize}
\item The CMB + 2dFGRS alone cannot rule out MDM; one needs 
a further prior, e.g. the Hubble constant from HST \cite{freedman}, 
supernovae Type Ia \cite{perlmutter,riess}, a prior on 
$\Omega_{\rm m}$ from the cluster baryon 
fraction \cite{pirin}, or evolution of cluster abundance 
with redshift \cite{lukash}.   
\item The statement that CMB + 2dFGRS alone provides evidence independent 
of that of supernovae Type Ia for a cosmological constant/dark energy 
is too strong.  This result is obtained only when neutrino masses 
are assumed to be negligible.   If one allows for 
massive neutrinos, acceptable fits to the CMB+2dFGRS can 
be obtained with $\Omega_{\rm m}=1$ and $\Omega_{\nu}\sim 0.2$.  
\end{itemize}

\subsection{The prior on the Hubble parameter} 

\begin{figure}
\begin{center}
\includegraphics[width=84mm]{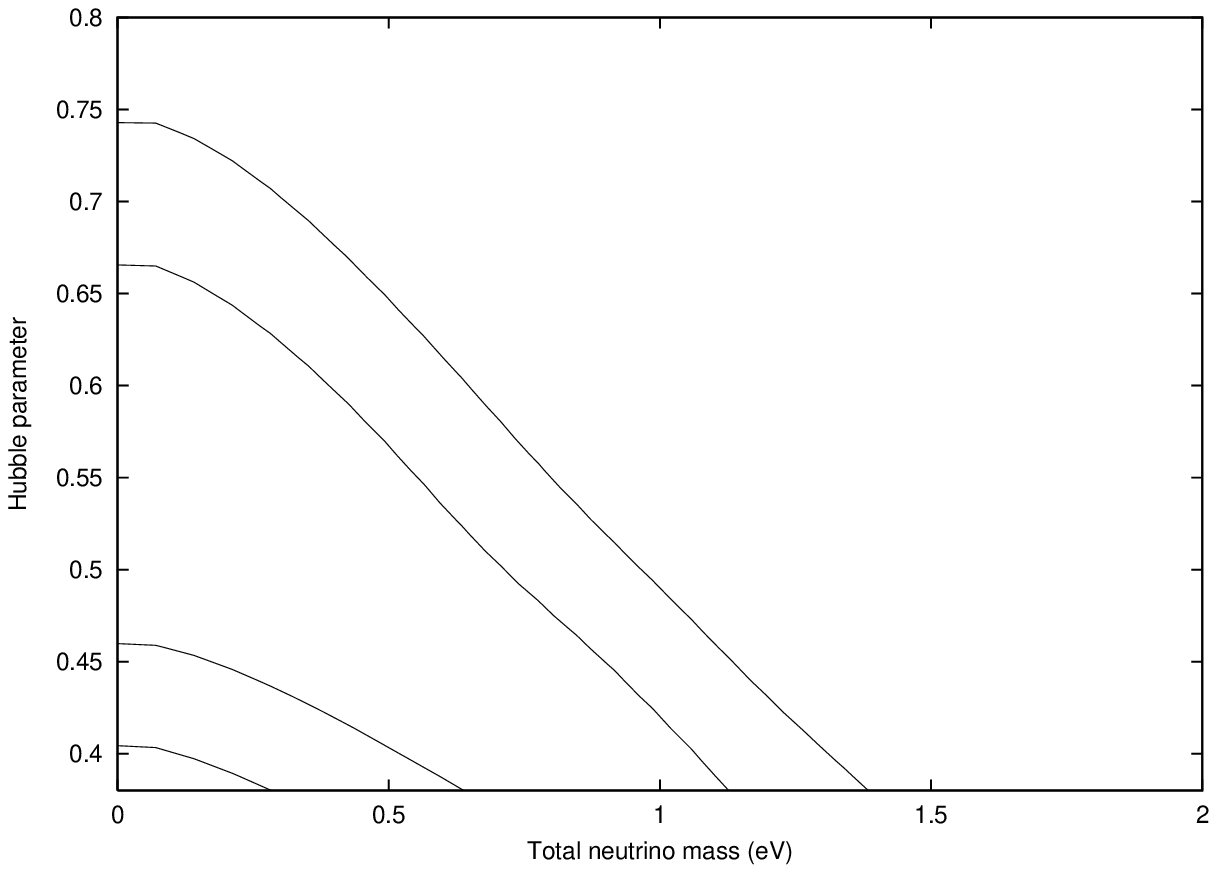}
\end{center}
\caption{Confidence contours (68 and 95 \%) from the 2dFGRS data alone 
in the $m_{\nu,\rm tot}$--
$h$ plane. The bias parameter and $\sigma_8$ have been marginalized over 
with top-hat priors, while $\omega_{\rm m}$, $\omega_{\rm b}$, and $n$ 
 have been fixed at their best-fitting values.}
\label{fig:fig7}
\end{figure}

We saw in the previous subsection that if we low values of the 
Hubble constant, $h<0.5$, MDM models provide reasonable fits 
to the CMB and 2dFGRS power spectra, but by combining with the 
HST prior on $h$, one will obtain the by now 
usual `concordance' values $\Omega_{\rm m}\sim 0.3$, $\Omega_{\rm \Lambda} 
\sim 0.7$.  The derivation of a strong upper limit on the total neutrino mass 
therefore  depends on our ability to exclude values of $h$ much below 0.7.   
Figure \ref{fig:fig7} shows the degeneracy between $h$ and $\omega_\nu$. 
If we allow $\Omega_{\rm m}=1$, and drop 
the prior on $h$, we find that a universe with a low value of $h<0.5$ 
is a viable option if the total neutrino mass is a few eVs.  
Note that these models also have ages $>12\;{\rm Gyr}$, consistent with 
ages of globular clusters \cite{chaboyer}.  Thus, without a prior on $h$, 
the CMB prior on $\Omega_{\rm m}$ 
would have been weaker, and which would also have affected the 
upper limit on the neutrino masses.  

\subsection{The prior on the scalar spectral index} 
\begin{figure}
\begin{center}
\includegraphics[width=84mm]{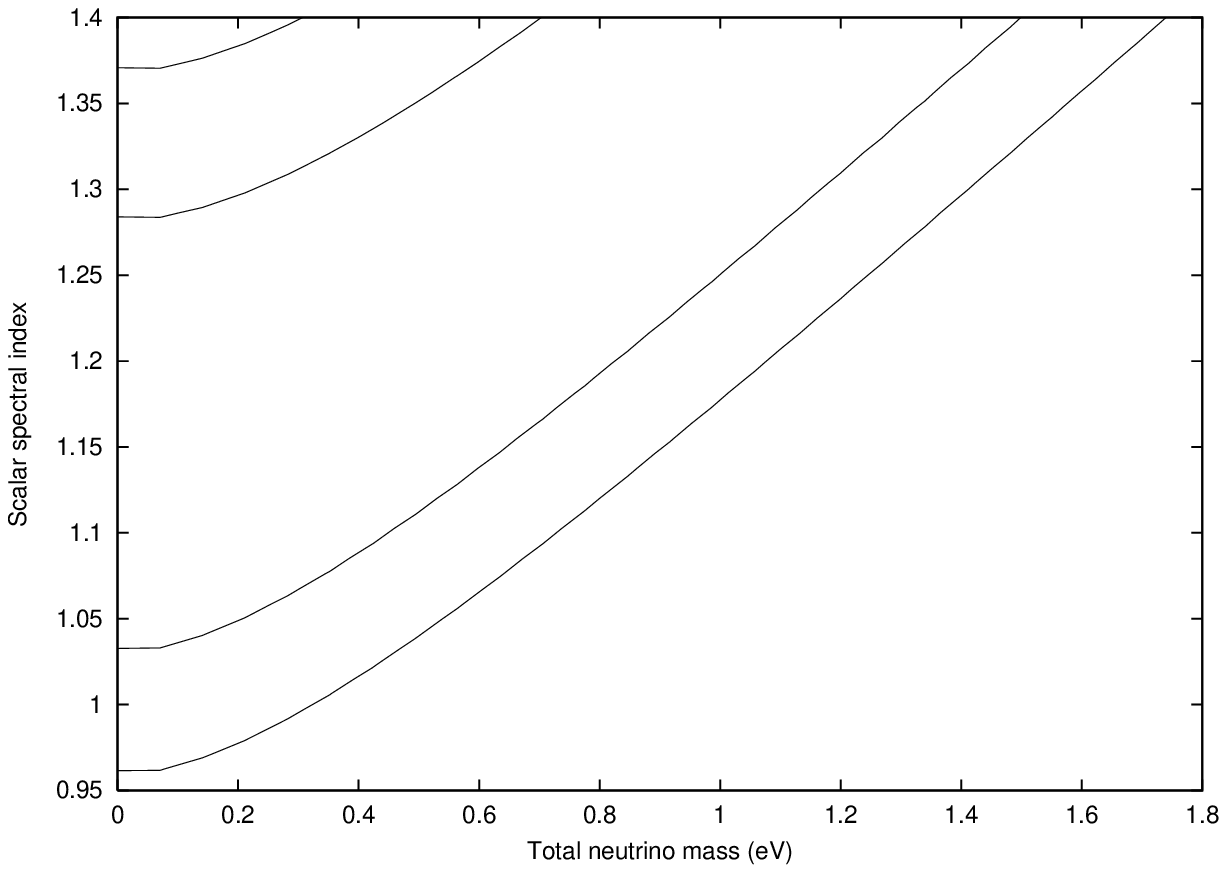}
\end{center}
\caption{Confidence contours (68 and 95 \%) from the 2dFGRS data alone 
in the $m_{\nu,\rm tot}$--$n$ plane.  The bias parameter and $\sigma_8$ 
have been marginalized over with top-hat priors, 
while $\omega_{\rm m}$, $\omega_{\rm b}$ and $h$  
have been fixed at their best-fitting values.  
}
\label{fig:fig8}
\end{figure}

As noted earlier, there is also a degeneracy between the scalar 
spectral index $n$ and $\omega_\nu$, illustrated in figure \ref{fig:fig8}.
Motivated by the pre-WMAP CMB data we used a prior $n=1.0\pm 0.1$ in 
\cite{elgar}.  For flat $\Lambda{\rm CDM}$ models the WMAP data 
give $n=0.99\pm 0.04$ (see table 1 in \cite{wmap}), but in their 
full analysis including other datasets there is some evidence for 
a running scalar spectral index.  However, it has been argued that 
this may be because of their treatment of the Lyman $\alpha$ forest 
power spectrum, and that in a more conservative approach one finds 
that a scale-invariant primordial power spectrum, $n=1$ is 
consistent with the data \cite{seljak2}.  We will therefore not 
consider a running spectral index in this paper.   

\subsection{The prior on the baryon density}
\begin{figure}
\begin{center}
\includegraphics[width=84mm]{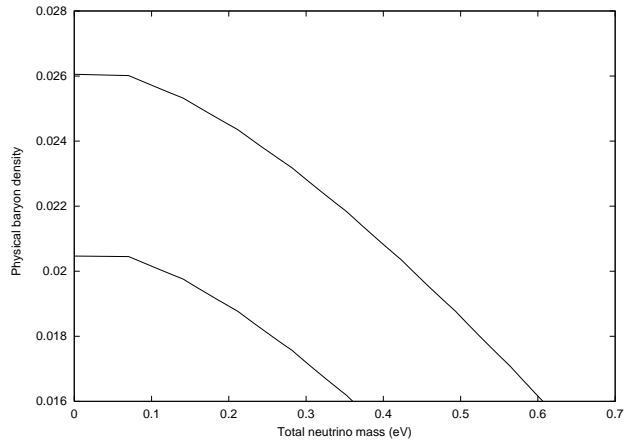}
\end{center}
\caption{Confidence contours (68 and 95 \%) from the 2dFGRS data alone 
in the $m_{\nu,\rm tot}$--$\omega_{\rm b}$ plane.  The bias parameter 
and $\sigma_8$ have been marginalized over with top-hat priors, while 
$\omega_{\rm m}$, $h$, and $n$ have been fixed at their best-fitting 
values. 
}
\label{fig:fig9}
\end{figure}

There is also some degeneracy between the baryon density and the 
neutrino masses, as shown in figure \ref{fig:fig9}, because increasing the 
baryon density suppresses power on small scales.  
The degeneracy is, however, less serious than for the other parameters, 
and WMAP provides a tight constraint on $\omega_{\rm b}$ 
from the ratio of the amplitudes of the first and second peaks in 
the CMB power spectrum, $\omega_{\rm b} = 0.024 \pm 0.001$, 
consistent with standard BBN \cite{burles}.

\subsection{Non-linear fluctuations} 

At small scales, one eventually enters the regime where the 
power spectrum is no longer linear.      
Therefore, as a further test of the stability of our analysis in 
\cite{elgar}, we used 
the full set of priors, but only the power spectrum data at 
scales $k < 0.1\;h\,{\rm Mpc}^{-1}$ and found that  
the limit increased to $f_\nu < 0.20$.  Cutting the power spectrum 
at this scale is, however, very conservative.  To see this, we 
follow the analysis of non-linearities in \cite{steen2}.  
Defining 
\begin{equation}
\Delta^2(k)\equiv \frac{k^2}{2\pi^2}P(k),
\label{eq:deltak}
\end{equation}
the crossover from linear to non-linear behaviour is at the co-moving 
momentum $k_{\rm cut}$ where $\Delta^2(k_{\rm cut}) = 1$.  This 
corresponds roughly to the point where $\Delta^2_{\rm non-linear}-
\Delta^2_{\rm linear} \sim \Delta^2_{\rm linear}$.  We use the approximate 
relation between the the linear and non-linear spectrum found by 
\cite{peadodds}.  
The cut in the spectrum should be made where the non-linear effects are of the 
same order as the suppression of small scale-power from massive 
neutrinos, which is given by equation (\ref{eq:fnusuppr}).  
For neutrino masses $\sim 1\;{\rm eV}$, this gives 
$\Delta^2_{\rm linear} \sim 3$.  We chose 
$k_{\rm cut}=0.15\;h\,{\rm Mpc}^{-1}$ in our analysis, and at that point 
$\Delta^2_{\rm linear}\approx 0.7$.  Crude as this argument may be, 
it clearly indicates that the neutrino mass signal dominates possible 
non-linear effects in the 2dFGRS power spectrum data used in our analysis.  
We note that the linear matter power spectrum is   
convolved with the survey window function before comparison with 
the data.  However, for values of $k$ greater than $\sim 0.1\;h\,{\rm Mpc}
^{-1}$ the window function is sharply peaked at $k$, and so there is 
little mixing with smaller scales \cite{gramann}. 
The WMAP team  went one step further and considered non-linear 
corrections at the smallest scales in the analysis.  Nevertheless, 
one would expect these effects to be small for $k < 0.15\;h\,{\rm Mpc}^{-1}$.

\subsection{Scale-dependent bias}

The power spectrum of the galaxy distribution
might be `biased' relative to the matter power spectrum,
hence it might introduce systematic error in the estimation
of the neutrino mass.

Indeed it is well established that on scales less than $\sim 10\;{\rm Mpc}$
different galaxy populations exhibit different clustering amplitudes,
the so-called morphology-density relation (e.g.
\cite{dressler,hermit,norberg,zehavi}). Hierarchical
merging scenarios also suggest a more complicated picture of biasing
as it could be non-linear, scale-dependent and stochastic (e.g.
\cite{benson,mowhite,matarrese,magliochetti,dekel,blanton,somerville}.
But is the biasing still scale-dependent at the large scales
($k^{-1} > 7\;h^{-1}\,{\rm Mpc}$) where we analyse the 2dFGRS
power spectrum ?  Let us consider the ratio of galaxy to matter power
spectra, and use the ratio of these to define the bias parameter 
as in equation (\ref{eq:biasfac}). 
To illustrate the dramatic effect that scale-dependent might have
we assume the following simple form:
\begin{equation}
b(k) = a\log(k/k^*)+c,
\label{eq:kbias}
\end{equation}
where we fix $k^*=0.15\;h\,{\rm Mpc}^{-1}$ (note that a shift in $k^*$
can be absorbed in a change in $c$), but allow  $a$ and $c$ to vary.

Analysis of the semi-analytic galaxy formation models
in \cite{berlind} shows that on large scales the biasing
function $b(k)$ is nearly constant to high degree.
In our parameterization (\ref{eq:kbias}) 
even the brightest galaxies  ($L>0.75 L_*$, where $L_*$ is the 
characteristic luminosity of the Schechter luminosity function) 
are weakly biased,
with slope $a <  0.15$ over the scales $ 3 <  k^{-1} <16  h^{-1} {\rm Mpc}$
\cite{berlindpriv}.   
The simulations in \cite{blanton} also suggest scale-independent
biasing on scales larger than 10 $h^{-1}\,{\rm Mpc}$ at late times.

Observationally, the bi-spectrum analysis of
the 2dFGRS showed almost no deviation from linear biasing \cite{verde}
and combined analysis of 2dFGRS with CMB data on scales of
$0.02 < k < 0.15 \;h\,{\rm Mpc}^{-1}$ \cite{lahav} gave
$b \sim 1$ for $L_*$ galaxies.
Furthermore, the ratio of the power-spectra of blue and red galaxies in
2dFGRS \cite{jap2003} 
is almost constant  over the range of our analysis,
$0.02 < k < 0.15 \;h\,{\rm Mpc}^{-1}$.
This suggests (as a necessary, but not sufficient condition) that the
galaxy power spectrum is proportional to the underlying matter power
spectrum.
Based on
these theoretical and observational studies we argue that
scale-dependent biasing is unlikely to pose a problem in estimating
the neutrino mass from the 2dFGRS.

\begin{figure}
\begin{center}
\includegraphics[width=84mm]{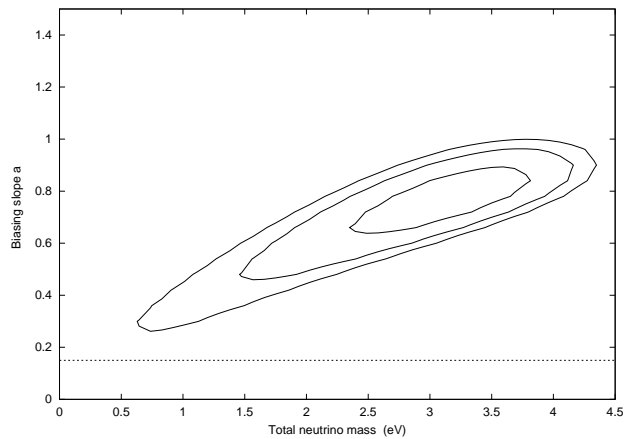}
\end{center}
\caption{Likelihood contours (68, 95, and 99 \%) in the 
$m_{\nu,\rm tot}$--$a$ plane after marginalizing over $c$, where 
$a$ and $c$ are defined in equation (\ref{eq:kbias}).   
The horizontal line $a=0.15$ is an 
upper limit estimated from simulations of biasing (see text).}
\label{fig:fig10}
\end{figure}

We then redo the analysis of the 2dFGRS $P_{\rm g}(k)$ with $\omega_\nu$,
$a$ and $c$ as free parameters.  The remaining parameters are fixed at
their `concordance' values ($\Omega_{\rm m}=0.3$, $h=0.7$ etc.).  
We distinguish
between two cases: $a \leq 0$ (bias increasing with length-scale
and $ a \geq 0$
(bias decreasing  with length-scale).  In the first case, the best fit is
found at
$\omega_\nu = 0$, $a=0$, whereas in the second case one finds the best
fit for $\omega_\nu \approx 0.029$, $a=0.72$, $c=1.1$, i.e. for a
non-zero neutrino mass $m_{\nu,\rm tot}\approx 2.8\;{\rm eV}$.
 This is understandable, since $b(k)$ in this
case has the opposite effect of massive neutrinos, so the two effects
can be `tuned' to give a very good fit to the 2dFGRS data with a
non-zero neutrino mass which is unrealistically high.
Figure \ref{fig:fig10}
shows the likelihood contours in the $m_{\nu,\rm tot}$-$a$ plane.
As argued above, simulations and observations argue
for $a < 0.15 $ at large scales, and in what follows we shall
assume constant biasing.

\subsection{The case of scale-independent bias in more detail}

WMAP provides tight constraints on 
$\omega_{\rm m}$, $h$, $n$, and $\omega_{\rm b}$, and we have seen 
in previous subsections that having good constraints on these 
parameters is essential to obtaining a good upper limit on 
$m_{\nu,\rm tot}$.   
Also, from the CMB one can constrain the amplitude of the matter power 
spectrum (quoted, e.g. in terms of the rms mass fluctuations in 
spheres of radius $8\;h^{-1}\,{\rm Mpc}$, $\sigma_8$).  
If the biasing and redshift-space distortions were known, 
then this would translate directly to 
a constraint on the amplitude of the 2dFGRS power spectrum, and 
could potentially tighten the constraint on $m_{\nu,\rm tot}$.    
The WMAP analysis makes use of the constraint on the 
amplitude of the matter power spectrum from the CMB by  
introducing a prior on the bias parameter, taking it to be a Gaussian 
with $b=1.04\pm 0.11$ \cite{wmapanalysis} as found in the analysis 
of the 2dFGRS bispectrum \cite{verde}.  As the analysis in \cite{verde} 
was performed for a different range of scales than those involved in the 
analysis of the linear part of the 2dFGRS power spectrum, and did not 
take neutrino masses into account (even though the cosmology dependence 
in the bispectrum analysis is mild) and questions have been raised 
about this approach \cite{raffelt}, it is worthwhile to take a closer 
look at the WMAP approach.  To do this, we need to go into the issue 
of a constant bias and redshift-space distortions  in more detail than in 
the previous sections, and we will 
do so following \cite{lahav}.  The WMAP analysis was more detailed 
\cite{wmapanalysis}, but we think that our simplified version 
captures the main points.  

We now need to take two effects into account: the fact that the 2dFGRS 
power spectrum is given in redshift space, and that the galaxy distribution 
may be biased with respect to the mass distribution.  At redshift $z=0$, 
the relation between the real-space normalization at redshift 0, 
$\sigma_{8\rm m}(0)$, of the matter power 
spectrum and the normalization of the galaxy power spectrum is given by 
\begin{equation}
\sigma_{8\rm g}^{\rm R}(L_s,0) = b(L_s,0)\sigma_{8\rm m}(0),
\label{eq:bias1}
\end{equation}
where $L_s\approx 1.9L_*$ for the 2dFGRS.  We follow \cite{lahav} 
and assume that galaxy clustering evolves weakly in the range of 
redshifts $0 < z < 0.2$ so that $\sigma_{8\rm g}^{\rm R}(L_s,z_s)\approx 
\sigma_{8\rm g}^{\rm R}(L_s,0)$, where $z_s \approx 0.17$.  
The conversion of $\sigma_{8\rm g}$ 
from real space to redshift space is determined by 
\begin{equation}
\sigma_{8 \rm g}^{\rm S}(L_s,z_s)=\sigma_{8\rm g}^{\rm R}(L_s,z_s)
\sqrt{K[\beta(L_s,z_s)]},
\label{eq:bias2}
\end{equation}
where 
\begin{equation}
K[\beta] = 1 + \frac{2}{3}\beta + \frac{1}{5}\beta^2,
\label{eq:bias3}
\end{equation}
is Kaiser's factor \cite{kaiser}, and 
\begin{equation}
\beta(L_s,z_s) \approx \frac{\Omega_{\rm m}^{0.6}(z_s)}{b(L_s,z_s)},
\label{eq:bias4}
\end{equation}  
with 
\begin{equation}
\Omega_{\rm m}(z) = \frac{\Omega_{\rm m}(1+z)^3}
{\Omega_{\rm m}(1+z)^3 + \Omega_\Lambda}.
\label{eq:bias5}
\end{equation}
Furthermore, we assume that the mass fluctuations grow as 
$\sigma_{8\rm m}(z) = \sigma_{8\rm m}(0)D(z)$, where $D(z)$ 
is the linear growth rate (normalized to 1 at $z=0$). 
As pointed out earlier, the linear growth rate is 
actually scale dependent in models with massive neutrinos.  
We have checked that this scale dependence is weak for the 
parameter range we consider, and evaluated $D(z)$ at 
$k=\overline{k}$, where $\overline{k}$ is the mean value of $k$ 
for a spherical top-hat window function.
With these assumptions we have 
\begin{equation}
b(L_s,z_s)=\frac{b(L_s,0)}{D(z_s)}.
\label{eq:bias6}
\end{equation}
Given $b(L_s,0)$, we can translate a given normalization $\sigma_{8\rm m}(0)$ 
of the matter power spectrum to $\sigma_{8\rm g}^{\rm S}(L_s,z_s)$ for 
the galaxy power spectrum.  

\begin{figure}
\begin{center}
\includegraphics[width=84mm]{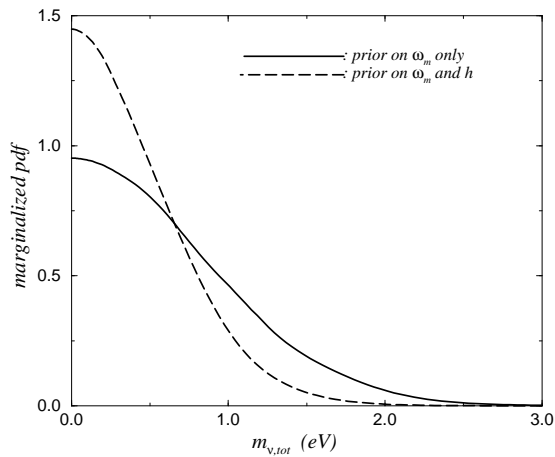}
\end{center}
\caption{Probability distributions for the total 
neutrino mass in our simplified analysis after marginalizing over 
$\omega_{\rm m}$, $h$, and $b$ for the cases of 
a prior on $\omega_{\rm m}$ only, and priors on $\omega_{\rm m}$ and $h$.  
When a Gaussian prior $b=1.04\pm 0.11$ is added, the resulting curve 
is indistinguishable from the dashed line.
}
\label{fig:fig11}
\end{figure}

\begin{figure}
\begin{center}
\includegraphics[width=84mm]{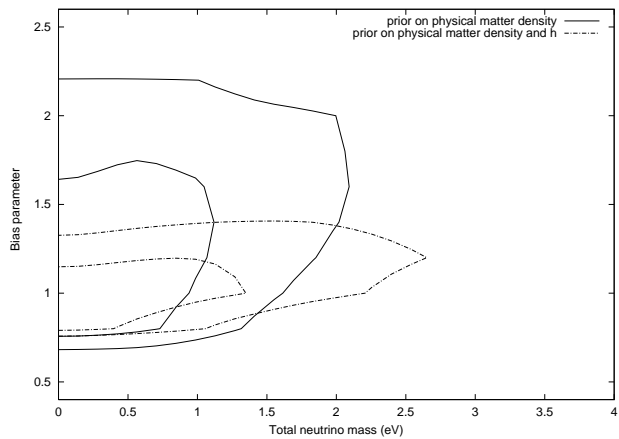}
\end{center}
\caption{Confidence contours (68 and 95 
plane after marginalizing over $\omega_{\rm m}$ and $h$ for the cases of 
a prior on $\omega_{\rm m}$, and priors on both $\omega_{\rm m}$ and $h$.  
}
\label{fig:fig12}
\end{figure}

We now carry out the following simple analysis: we fix $n=0.99$, 
$\Omega_{\rm b}h^2 = 0.024$ (all values taken from table 
1 in \cite{wmap} for WMAP alone), 
and fit $\omega_\nu$, $\omega_{\rm m}$, $h$, $\sigma_{8\rm m}$ and 
$b(L_s,0)$ to the combination of the 2dFGRS power spectrum data and 
the  constraint $\sigma_{8\rm m}\Omega_{\rm m}^{0.6}=0.44\pm 0.10$ 
from table 2 in \cite{wmap}.    
Furthermore, we add Gaussian priors $\Omega_{\rm m}h^2 = 0.14\pm 0.02$, 
$h = 0.72\pm 0.05$ 
from WMAP, and look at the limit on $m_{\nu,\rm tot} = 94\Omega_\nu 
h^2\;{\rm eV}$ for the cases of with and without a prior on 
$b\equiv b(L_s,0)$.  The results reveal that the prior on $\omega_{\rm m}$  
is crucial and the prior on $h$ very important, but also that, 
at least in this 
simplified analysis, the effect of adding a prior on the bias is 
very small, as can be seen from figure \ref{fig:fig11}.  
Without a prior on $\omega_{\rm m}$, no non-trivial limit on 
$m_{\nu,\rm tot}$ is obtained.  With just the $\omega_{\rm m}$ prior, 
the 95 \% confidence limit is $m_{\nu,\rm tot} < 1.6\;{\rm eV}$, 
and this improves to $1.1\;{\rm eV}$ when the prior on $h$ is added.   
Adding the prior on $b$ does not change the limit on $m_{\nu,\rm tot}$, 
it stays at $1.1\;{\rm eV}$.  From the contour plots in 
figure \ref{fig:fig12}, 
the degeneracy between $\omega_{\nu}$ and $b$ is small, especially 
when the prior on $\omega_{\rm m}$ is included. 
This is not 
in contradiction to the analysis in \cite{steen3}, because 
our treatment of biasing is different: we have treated the 
redshift-space distortions explicitly, and then the constraint 
on $\sigma_8$ and $\omega_{\rm m}$ from the CMB breaks much of 
the degeneracy between $\omega_\nu$ and $b$.  This is because 
the redshift distortion itself depends on $\Omega_{\rm m}$, 
see equations (\ref{eq:bias3},\ref{eq:bias4}).

In this simple analysis we get a 95 \% confidence limit 
of $m_{\nu,\rm tot} < 1.1 \;{\rm eV}$.  This is still 
some way from the WMAP limit of 0.71 eV, even with our very 
restricted parameter 
space, but consistent with the analysis in \cite{steen3}.   
The WMAP analysis also used data from ACBAR and CBI \cite{acbar,cbi}, 
and included the Lyman $\alpha$ 
forest power spectrum.  The linear matter power spectrum inferred from  
the Lyman $\alpha$ fores probes smaller scales than the 2dFGRS 
and therefore has considerable power in constraining neutrino masses.   
We have seen that the most severe degeneracies of $\omega_\nu$ 
are with $\omega_{\rm m}$, $n$, $h$, and $\omega_{\rm b}$.  
The most serious one is with $\omega_{\rm m}$: 
without a prior on the physical matter density, one cannot get a 
non-trivial upper bound on $m_{\nu,\rm tot}$.  This makes sense, as 
the matter power spectrum depends on $f_\nu = \Omega_\nu / \Omega_{\rm m}$ 
and this is why the analysis was carried out in terms of this parameter 
in \cite{elgar}.  So the fact that WMAP (restricted to flat $\Lambda{\rm CDM}$ 
models and $h > 0.5$) provides a tight constraint $\omega_{\rm m} = 0.14 \pm 0.02$ is crucial for constraining neutrino masses.  WMAP also constrains 
the spectral index to a narrow interval around $n=1$, and, just as 
importantly, by constraining $h$ to be around 0.7 eliminates the 
possibility for MDM models with large neutrino masses to give good 
fits to the 2dFGRS power spectrum.  The importance of the prior on 
$h$ was also noted in \cite{steen3}.

\section{Other cosmological probes of neutrino masses}

Direct probes of the 
total matter distribution avoid the issue of biasing and are therefore 
ideally suited for providing limits on the neutrino masses. 
Several ideas for how this can be done exist.  In \cite{fukugita} 
the normalization 
of the matter power spectrum on large scales derived from COBE 
was combined with constraints on $\sigma_8$ from cluster abundances 
and a constraint $m_{\nu,\rm tot} < 2.7\;{\rm eV}$ obtained, although 
with a fairly restricted parameter space.  However, $\sigma_8$ is 
probably one of the most debated numbers in cosmology at the moment 
\cite{wang},  
and a better understanding of systematic uncertainties connected with 
the various methods for extracting it from observations is needed 
before this method can provide useful constraints.  The potential 
of this method to push the value of the mass limit down also depends 
on the actual value of $\sigma_{\rm 8}$: the higher $\sigma_8$ turns 
out to be, the less room there will be for massive neutrinos.  
The evolution of cluster abundance with redshift may  provide  
further constraints on neutrino masses \cite{lukash}.  
The Lyman $\alpha$ forest provides constraints on the matter power spectrum 
on small scales, where the effect of massive neutrinos is most visible, 
and it was used in \cite{croft} to derive  
a limit $m_{\nu,\rm tot} < 5.5\;{\rm eV}$, and it clearly played a role 
in the WMAP limit also.  How to use this probe correctly in cosmological 
parameter estimation is, however, still being discussed \cite{seljak2}.  
Massive neutrinos also suppress peculiar velocities 
on scales smaller than $50\;h^{-1}\,{\rm Mpc}$,  
where they can be measured more 
accurately to nearby galaxies, however, non-linear effects on small 
scales causes complications.  Finally, deep and wide weak lensing 
surveys will in the future make it possible to do weak lensing tomography 
of the matter density field \cite{hulens1,hulens2}, and in 
\cite{dodelson} it has been shown that one can probe neutrino masses 
below $0.1\;{\rm eV}$ in this way.  However, this is under the 
assumption that the equation of state of the dark energy is known.

\section{Conclusions}

We have reviewed how the constraint on the neutrino mass in \cite{elgar}
was obtained, and the recent improved limit from WMAP, paying
attention to issues of priors and parameter degeneracies.
We have seen that one can derive fairly tight constraints
on neutrino masses from the 2dFGRS power spectrum, provided that one
has good constraints on $\omega_{\rm m}$, $n$, $h$, and $\omega_{\rm b}$
from independent data sets.

We found that external constraints on the Hubble parameter, for example the 
HST Key project,  is important in 
order to get a strong limit on the neutrino mass since, intriguingly,
the out-of-fashion Mixed Dark
Matter (MDM) model can give a reasonable description of the CMB and
2dFGRS data with $\Omega_\nu=0.2$, $\Omega_{\rm m}=1$ and no
cosmological constant if we allow low values of the Hubble parameter,
$h<0.5$.  The importance of having a prior on $h$ was also noted in \cite{steen3}.
We note that the above MDM model is inconsistent with other cosmic
measurements such as Supernovae Type Ia, baryon
fraction in clusters, and the evolution of cluster abundance with redshift, so 
adding any one of these datasets to the analysis would eliminate MDM and improve 
the neutrino mass limit.

We also considered the effect of
the possible bias of the galaxy distribution with respect to the
mass distribution on the neutrino mass limit.  A scale-dependent
bias has serious implications for the $m_{\nu,\rm tot}$ constraint,
but based on semi-analytic galaxy formation models \cite{berlind} the
scale-dependence of the bias is expected to be too weak to be of
any major concern on the large scales used in the analysis of the
2dFGRS power spectrum.
When the effects of redshift-space distortions on the 2dFGRS power
spectrum are included in the analysis, there is almost
no degeneracy between a constant, scale-independent bias factor
and the neutrino mass.  However, in our restricted analysis
we did not get as good a neutrino mass constraint with
2dFGRS + WMAP priors as in the full analysis in \cite{wmap} which
suggests that the Lyman $\alpha$ forest power spectrum plays a
role in pushing the constraint on $m_{\nu,\rm tot}$ below 1 eV.

\ack
We thank the 2dFGRS team, and in particular 
John Peacock for drawing our attention to the MDM issue and for 
useful comments.  
We acknowledge fruitful discussions 
with the Leverhulme Quantitative Cosmology Group in Cambridge.  
We also thank Andreas Berlind, David Weinberg, and Steen Hannestad for 
very helpful comments. 
\O E and OL wish to thank, respectively, the IoA in Cambridge and NORDITA 
for their hospitality.


\section*{References}
\begin{thebibliography}{99}
\bibitem{wmap}
Spergel D N, Verde L, Peiris H V, Komatsu E, Nolta M R, Bennett C L, 
Halpern M, Hinshaw G, Jarosik N, Kogut A, Limon M, Meyer S S, Page L, 
Tucker G S, Weiland J L , Wollack E, and Wright E L,  
{\it First Year Wilkinson Microwave Anisotropy Probe (WMAP) Observations: 
Determination of Cosmological Parameters}, 2003 {\it Preprint} astro-ph/0302209 

\bibitem{sage} 
Abdurashitov J N {\it et al.}, {\it Measurement of the solar neutrino capture rate 
with gallium metal}, 1999 {\it Phys. Rev. C} {\bf 60} 055801 [astro-ph/9907113]

\bibitem{sno}
Ahmad Q R {\it et al.}, {\it Measurement of the rate of $\nu_{\rm e}+{\rm d}
\rightarrow {\rm p}+{\rm p}+{\rm e}^{-}$ interactions produced by $^8{\rm B}$ 
solar neutrinos at the Sudbury Neutrino Observatory}, 2001 {\it Phys. Rev. Lett.} {\bf 87} 
071301 [nucl-ex/0106015] 


\bibitem{macro} 
Ambrosia M {\it et al.}, {\it Matter effects in upward-going muons and sterile 
neutrino oscillations}, 2001 {\it Phys. Lett.} B {\bf 517} 59 [hep-ex/0106049] 

\bibitem{gno} 
Altmann M {\it et al.}, {\it GNO solar neutrino observations: results for GNO I}, 
2000 {\it Phys. Lett.} B {\bf 490} 16 [hep-ex/0006034] 

\bibitem{homestake}
Cleveland B T {\it et al.}, {\it Measurement of the solar electron neutrino 
flux with the Homestake Chlorine Detector}, 1998 {\it Astrophys. J.} {\bf 496} 
505 

\bibitem{superk}
Fukuda S {\it et al.}, {\it Tau neutrinos favored over sterile neutrinos in 
athmospheric muon neutrino oscillations}, 2000 {\it Phys. Rev. Lett.} {\bf 85} 3999 


\bibitem{gallex}
{\it The Gallex collaboration}, {\it GALLEX solar neutrino observations: results 
for GALLEX IV}. 1999 {\it Phys. Lett.} B {\bf 447} 127 

\bibitem{primack}
Primack J R and Gross M A K, {\it Hot Dark Matter in Cosmology}, 2000 {\it Current Aspects 
of Neutrino Physics} (Berlin: Springer) p 287 [astro-ph/0007165]

\bibitem{mainz} 
Bonn J {\it et al.}, {\it The Mainz Neutrino Mass Experiment}, 2001 {\it 
Nucl. Phys. Proc. Suppl.} {\bf 91} 273 


\bibitem{klapdor1}
Klapdor-Kleingrothaus H V {\it et al.}, {\it Latest results from the HEIDELBERG-MOSCOW 
double beta decay experiment}, 2001 {\it Eur. Phys. J.} A {\bf 12} 147 
[hep-ph/0103062]


\bibitem{elgar} 
Elgar\o y \O. {\it et al.} (the 2dFGRS team), {\it New upper limit on the 
total neutrino mass from the 2 degree Field Galaxy Redshift Survey}, 
2002 {\it Phys. Rev. Lett.} {\bf 89} 061301 [astro-ph/0204152]

\bibitem{lewis}
Lewis A and Bridle S L, {\it Cosmological parameters from CMB and other data: 
A Monte Carlo approach}, 2002 {\it Phys. Rev.} D {\bf 66} 103511 
[astro-ph/0205436] 

\bibitem{acbar}
Kuo C L, Ade P A R, Bock J J, Cantalupo C,  Daub M D, 
Goldstein J, Holzapfel W L, Lange A E, Lueker M, Newcomb M, 
Peterson J B, Ruhl J, Runyan M C and Torbet E 2002,
{\it High resolution observations of the CMB power spectrum with 
ACBAR}, 2002 {\it Preprint} astro-ph/0212289 

\bibitem{cbi}
Pearson T J {\it et al.} (the CBI team), 
{\it The anisotropy of the microwave background to $\ell=3500$: 
Mosaic observations with the Cosmic Microwave Background Imager}, 
2002 {\it Preprint} astro-ph/0205388

\bibitem{lyalpha1}
Croft R A C, Weinberg D H, Bolte M, Burles S, 
Hernquist L, Katz N, Kirkman D and Tytler D, {\it Toward a precise measurement 
of matter clustering: Ly $\alpha$ forest data at redshifts 2-4}, 2002  
{\it Astrophys. J.} {\bf 581} 20 [astro-ph/0012324]  

\bibitem{gnedin}
Gnedin N Y and Hamilton A J S, {\it Matter power spectrum from the Lyman-alpha forest: 
myth or reality ?}, 2002 {\it Mon. Not. R. Astron. Soc.}  {\bf 334} 107 
[astro-ph/0111194] 
 

\bibitem{SDSS}
http://www.sdss.org

\bibitem{hutegmark}
Hu W, Eisenstein D and Tegmark M, {\it Weighing neutrinos with galaxy surveys}, 1998 
{\it Phys. Rev. Lett.} {\bf 80} 5255 [astro-ph/9712057]


\bibitem{LSND}
Aguilar A {\it et al.} (the LSND collaboration), 
{\it Evidence for neutrino oscillations from the observation of 
$\overline{\nu}_{\rm e}$ appearance in a $\overline{\nu}_{\mu}$ beam}, 
2001 {\it Phys. Rev.} D {\bf 64} 112007 [hep-ex/0104049]


\bibitem{pierce}
Pierce A and Murayama H, {\it WMAPping out neutrino masses}, 2003 
{\it Preprint} hep-ph/0302131 

\bibitem{bhatta}
Bhattacharyya G, P\"{a}s H, Song L and Weiler T J, 
{\it Particle physics implications of the WMAP neutrino mass bound}, 2003 
{\it Preprint} hep-ph/0302191.

\bibitem{steen3}
Hannestad S, {\it Neutrino masses and the number of neutrino species 
from WMAP and 2dFGRS}, 2003 {\it Preprint} astro-ph/0303076


\bibitem{efstathiou}
Efstathiou G P {\it et al.} (the 2dFGRS team), 
{\it Evidence for a non-zero $\Lambda$ and a low matter density from a 
combined analysis of the 2dF Galaxy Redshift Survey and cosmic microwave 
background anisotropies}, 2002  
{\it Mon. Not. R. Astron. Soc.}  {\bf 330} L29 [astro-ph/0109152] 


\bibitem{eguchi}
Eguchi K {\it et al.} (the KamLAND collaboration), 
{\it First results from KamLAND: Evidence for Antineutrino Disappearance}, 
2003 Phys. Rev. Lett. {\bf 90} 021802 [hep-ex/0212021] 

\bibitem{dolgov}
Dolgov A D, Hansen S H, Pastor S, Petcov S T, Raffelt G G  
and Semikoz D V, 
{\it Cosmological bounds on neutrino degeneracy improved by 
flavor oscillations}, 2002 {\it Nucl. Phys.} B {\bf 632} 363 
[hep-ph/0201287] 

\bibitem{ywong} Wong Y Y Y, 
{\it Analytical treatment of neutrino asymmetry equilibration from flavor 
oscillations in the early universe}, 2002 {\it Phys. Rev.} D {\bf 66} 025015 
[hep-ph/0203180]

\bibitem{kevork}
Abazajian K N, Beacom J F and Bell N F,
{\it Stringent constraints on cosmological neutrino-antineutrino 
asymmetries from synchronized flavor transformation}, 2002 
{\it Phys. Rev.} D {\bf 66}, 013008 [astro-ph/0203442]


\bibitem{elena}
Pierpaoli E, {\it Constraints on the cosmic neutrino background}, 
2003 {\it Preprint} astro-ph/0302465 

\bibitem{pakvasa}
Pakvasa S and Valle J W F, 
{\it Neutrino properties before and after KamLAND}, 
2003 {\it Preprint} hep-ph/0301061 


\bibitem{mboone}
Bazarko A {\it et al.} (the MiniBooNE collaboration), 
{\it MiniBooNE: Status of the Booster Neutrino Experiment}, 2000  
{\it Nucl. Phys. B. Proc. Suppl.} {\bf 91} 210 [hep-ex/0009056] 

\bibitem{eisenstein}
Eisenstein D J and Hu W, 
{\it Power spectra for Cold Dark Matter and its variants}, 1999 
{\it Astrophys. J.} {\bf 511} 5 [astro-ph/9710252]

\bibitem{zaldarriaga}
Seljak U and Zaldarriaga M, 
{\it A line-of-sight integration approach to cosmic microwave background 
anisotropies}, 1996 {\it Astrophys. J.} {\bf 469} 437 [astro-ph/9603033]

\bibitem{camb}
Lewis A, Challinor A and Lasenby A, 
{\it Efficient computation of cosmic microwave background anisotropies in 
closed Friedmann-Robertson-Walker models}, 2000  
{\it Astrophys. J.} {\bf 538}, 473 [astro-ph/9911177]

\bibitem{colless}
Colless M {\it et al.} (the 2dFGRS team), 
{\it The 2dF Galaxy Redshift Survey: spectra and redshifts}, 2001 
{\it Mon. Not. R. Astron. Soc.}  {\bf 328} 1039 [astro-ph/0106498]

\bibitem{percival}
Percival W J {\it et al.} (the 2dFGRS team), 
{\it The 2dF Galaxy Redshift Survey: The power spectrum and the matter content 
of the universe}, 2001 {\it Mon. Not. R. Astron. Soc.} {\bf 327} 1297 [astro-ph/0105252]

\bibitem{gramann}
Elgar\o y \O, Gramann M and Lahav O,
{\it Features in the primordial power spectrum: constraints from the cosmic microwave 
background and the limitation of the 2dF and SDSS redshift surveys to detect them}, 
2002 {\it Mon. Not. R. Astron. Soc.}  {\bf 333} 93 [astro-ph/0111208] 


\bibitem{benson}
Benson A J, Cole S, Frenk C S, Baugh C M and Lacey C G, 
{\it The nature of galaxy bias and clustering}, 
2000 {\it Mon. Not. R. Astron. Soc.}  {\bf 311} 793 [astro-ph/9903343]

\bibitem{lahav} 
Lahav O {\it et al.} (the 2dFGRS team), 
{\it The 2dF Galaxy Redshift Survey: the amplitudes of fluctuations in the 
2dFGRS and the CMB, and implications for galaxy biasing}, 2002 
{\it Mon. Not. R. Astron. Soc.}  {\bf 333} 961 [astro-ph/0112162]

\bibitem{verde}
Verde L {\it et al} (the 2dFGRS team), 
{\it The 2dF Galaxy Redshift Survey: the bias of galaxies and the density of 
the Universe}, 2002 
{\it Mon. Not. R. Astron. Soc.} {\bf 335} 432 [astro-ph/0112161]

\bibitem{freedman} 
Freedman W L {\it et al.}, 
{\it Final results from the Hubble Space Telescope Key Project to measure 
the Hubble constant}, 2001 {\it Astrophys. J} {\bf 553} 47 [astro-ph/0012376]

\bibitem{burles} 
Burles S, Nollett K M and Turner M S,
{\it What is the big-bang-nucleosynthesis prediction for the baryon density 
and how reliable is it ?}, 2001 {\it Phys. Rev.} D {\bf 63} 063512 
[astro-ph/0008495]



\bibitem{perlmutter}
Perlmutter S {\it et al.}, {\it Measurements of Omega and Lambda from 42 
high-redshift supernovae}, 
1999 {\it Astrophys. J.} {\bf 517} 565 [astro-ph/9812133] 

\bibitem{riess}
Riess A G {\it et al},
{\it BVRI light curves for 22 Type IA supernovae}, 
1999 {\it Astronom. J.} {\bf 117} 707 [astro-ph/9810291]

\bibitem{wang}
Wang X, Tegmark M, Jain B and Zaldarriaga M, 
{\it The last stand before MAP: cosmological parameters from lensing, CMB and 
galaxy clustering}, 2002 {\it Preprint} astro-ph/0212417 


\bibitem{lukash}
Arhipova N A, Kahniashvili T and Lukash V N, 
{\it Abundance and evolution of galaxy clusters in cosmological models with 
massive neutrino}, 2002 {\it Astron. Astrophys.} {\bf 386} 775 [astro-ph/0110426]


\bibitem{pirin}
Erdogdu P, Ettori S and Lahav O,
{\it `Hyper Parameters' approach to joint estimation: applications to 
Cepheid-calibrated distances and X-ray clusters}, 2003  
{\it Mon. Not. R. Astron. Soc.} in press [astro-ph/0202357]

\bibitem{chaboyer} 
Krauss L M and Chaboyer B, 
{\it Age estimates of globular clusters in the Milky Way: constraints on cosmology}, 
2003 {\it Science} {\bf 299} 65  


\bibitem{seljak2}
Seljak U, McDonald P and Makarov A,
{\it Cosmological constraints from the CMB and Ly-alpha forest revisited}, 
2003 {\it Preprint} astro-ph/0302571 


\bibitem{steen2}
Hannestad S, 
{\it Can cosmology detect hierarchical neutrino masses ?}, 2002 
{\it Preprint} astro-ph/0211106 



\bibitem{peadodds} 
Peacock J A and Dodds S J, 
{\it Reconstructing the linear power spectrum of cosmological mass fluctuations}, 
1994 {\it Mon. Not. R. Astron. Soc.}  {\bf 267} 1020 [astro-ph/9311057]



\bibitem{dressler}
Dressler A, 
{\it Galaxy morphology in rich clusters-Implications for the formation and evolution 
of galaxies}, 1980 {\it Astrophys. J.} {\bf 236} 351  

\bibitem{hermit} 
Hermit S, Santiago B X, Lahav O, Strauss M A, Davis M, 
Dressler A  and Huchra J P, 
{\it The two-point correlation function and morphological segregation in 
the Optical Redshift Survey}, 1996 {\it Mon. Not. R. Astron. Soc.}  {\bf 283}  
709 [astro-ph/9608001] 

\bibitem{norberg}
Norberg P {\it et al.} (the 2dFGRS team), 
{\it The 2dF Galaxy Redshift Survey: the dependence of galaxy clustering on 
luminosity and spectral type}, 2002 {\it Mon. Not. R. Astron. Soc.}  {\bf 332} 827 
[astro-ph/0112043]

\bibitem{zehavi}
Zehavi I {\it et al.} (SDSS Collaboration), 
{\it Galaxy clustering in early Sloan Digital Sky Survey Redshift data}, 
2002 {\it Astrophys. J.} {\bf 571} 172 [astro-ph/0106476] 

\bibitem{mowhite}
Mo H J and White S D M, 
{\it An analytic model for the spatial clustering of dark matter haloes}, 
1996 {\it Mon. Not. R. Astron. Soc.}  {\bf 282} 347 [astro-ph/9512127]


\bibitem{matarrese}
Matarrese S, Coles P, Lucchin F and Moscardini L, 
{\it Redshift evolution of clustering}, 1997  
{\it Mon. Not. R. Astron. Soc.}  {\bf 286} 115 [astro-ph/9608004] 

\bibitem{magliochetti}
Magliochetti M, Bagla J, Maddox S J and Lahav O, 
{\it The observed evolution of galaxy clustering vs. epoch-dependent biasing 
models}, 2000 {\it Mon. Not. R. Astron. Soc.}  {\bf 314} 546  
[astro-ph/9902260] 

\bibitem{dekel}
Dekel A and Lahav O, 
{\it Stochastic nonlinear galaxy biasing}, 
1999 {\it Astrophys. J.} {\bf 520} 24 [astro-ph/9806193]

\bibitem{blanton}
Blanton M, Cen R, Ostriker J P, Strauss M A and Tegmark M, 
{\it Time evolution of galaxy formatino and bias in cosmological simulations}, 
2000 {\it Astrophys. J.} {\bf 531} 1 [astro-ph/9903165]

\bibitem{somerville}
Somerville R, Lemson G, Sigad Y, Dekel A, Colberg J, Kauffmann G  
and White S D M,
{\it Non-linear stochastic galaxy biasing in cosmological simulations}, 
2001 {\it Mon. Not. R. Astron. Soc.} {\bf 320} 289 [astro-ph/9912073]

\bibitem{berlind}
Berlind A A, Weinberg D H, Benson A J, Baugh C M, Cole S, 
Dav\'{e} R, Frenk C S, Katz N and Lacey C G, 
{\it The halo occupation distribution and the physics of galaxy formation}, 
2002 {\it Preprint} astro-ph/0212357 

\bibitem{berlindpriv}
Berlind A A, private communication; the models themselves are 
similar to those in \cite{benson}. 


\bibitem{jap2003}
Peacock J A, 
{\it Implications of 2dFGRS results on cosmic structure}, 
2003 {\it Preprint} astro-ph/0301042 

\bibitem{wmapanalysis}
Verde L {\it et al.} (the WMAP team), 
{\it First year Wilkinson Microwave Anisotropy Probe (WMAP) observations: 
parameter estimatino methodology}, 
2003 {\it Preprint} astro-ph/0302218 

\bibitem{raffelt}
G. G. Raffelt, 
{\it Neutrinos in physics and astrophysics}, 2003 {\it Preprint} astro-ph/0302589 

\bibitem{kaiser}
Kaiser N, 
{\it Clustering in real space and in redshift space}, 1987 
{\it Mon. Not. R. Astron. Soc.}  {\bf 227} 1 


\bibitem{fukugita}
Fukugita M, Liu G-C and Sugiyama N, 
{\it Limits on neutrino mass from cosmic structure formation},  
2000 {\it Phys. Rev. Lett.} {\bf 84} 1082 [astro-ph/9908450] 

\bibitem{croft}
Croft R A C, Hu W and Dav\'{e} R, 
{\it Cosmological limits on the neutrino mass from the LyAlpha forest}, 
1999 {\it Phys. Rev. Lett.} {\bf 83} 1092 [astro-ph/9903335]


\bibitem{hulens1}
Hu W, 
{\it Power spectrum tomography with weak lensing}, 
1999 {\it Astrophys. J.} {\bf 522} 21 [astro-ph/9904152]
\bibitem{hulens2}
Hu W, 
{\it Dark energy and matter evolution from lensing tomography}, 
2002 {\it Phys. Rev.} D {\bf 66} 083515 [astro-ph/0208093] 
\bibitem{dodelson}
Abazajian K N and Dodelson S, 
{\it Neutrino mass and dark energy from weak lensing}, 
2002 {\it Preprint} astro-ph/0212216  


\end{thebibliography}
\end{document}